\documentclass[final,onecolumn,12pt]{IEEEtran}
\usepackage{graphicx,psfrag,epsfig,epsf,latexsym,hhline,amsmath,amssymb,multirow}
\usepackage[usenames,dvipsnames]{pstricks}
\usepackage{pst-plot}
\usepackage{pstricks-add}
\usepackage{hyperref}
\usepackage{graphicx}
\usepackage{amsthm}
\usepackage{algorithm}
\usepackage{algpseudocode}
\usepackage[noadjust]{cite}
\usepackage{blindtext}
\usepackage{etoolbox}

\newcommand{\mc}[1]{\mathcal{#1}}

\newcommand{\mbb}[1]{\mathbb{#1}}

\newcommand{\msf}[1]{\mathsf{#1}}
\newcommand{\mfk}[1]{\mathfrak{#1}}

\newcommand{\defeq}{\triangleq}
\newcommand{\Pp}{\mathbb{P}}

\newcommand{\Zw}{\mathbb{Z}[\omega]}
\newcommand{\Zi}{\mathbb{Z}[i]}

\newcommand{\Qd}{\mathbb{Q}(\sqrt{d})}
\newcommand{\Ok}{\mfk{O}_{\mbb{K}}}
\newcommand{\blambda}{\boldsymbol\lambda}

\newcommand{\iid}{i.\@i.\@d.\ }

\DeclareMathOperator{\var}{var}

\theoremstyle{definition}\newtheorem{lemma}{Lemma}
\theoremstyle{definition}
\theoremstyle{definition}\newtheorem{theorem}[lemma]{Theorem}
\theoremstyle{definition}
\newtheorem{define}[lemma]{Definition}
\newtheorem{example}[lemma]{Example}

\newtheorem{remark}[lemma]{Remark}

\newcommand\shortintertext[1]{%
\ifvmode\else\\\@empty\fi
\noalign{%
\penalty0%
\vbox{\mathstrut}%
\penalty10000%
\vskip-\baselineskip
\penalty10000%
\vbox to 0pt{%
\normalbaselines
\ifdim\linewidth=\columnwidth
\else
\parshape\@ne
\@totalleftmargin\linewidth
\fi
\vss
\noindent#1\par}%
\penalty10000%
\vskip-\baselineskip}%
\penalty10000}

\begin{document}
\title{Adaptive Compute-and-Forward with Lattice Codes Over Algebraic Integers\footnote{This paper was published in part at the 2015 International Symposium on Information Theory \cite{huang15ACF}.}}
\author{\S Yu-Chih Huang, \dag Krishna R. Narayanan, and \dag Ping-Chung Wang \\
\S Department of Communication Engineering, National Taipei University\\
\dag Department of Electrical and Computer Engineering, Texas A\&M University\\
{\tt\small {\{ychuang@mail.ntpu.edu.tw, krn@ece.tamu.edu, michael422603@tamu.edu\}} }
\thanks{The work of Y.-C. Huang was supported by the Ministry of Science and Technology, Taiwan, under Grant MOST 104-2218-E-305-001-MY2. The work of K. R. Narayanan and P.-C. Wang was funded in part by the National Science Foundation under Grant CCF 1302616. }}

\maketitle

\begin{abstract}
We consider the compute-and-forward paradigm with limited feedback. Without feedback, compute-and-forward is typically realized with lattice codes over the ring of integers, the ring of Gaussian integers, or the ring of Eisenstein integers, which are all principal ideal domains (PID). A novel scheme called adaptive compute-and-forward is proposed to exploit the limited feedback about the channel state by working with the best ring of imaginary quadratic integers. This is enabled by generalizing the famous Construction A from PID to other rings of imaginary quadratic integers which may not form PID and by showing such the construction can produce good lattices for coding in the sense of Poltyrev and for MSE quantization. Simulation results show that by adaptively choosing the best ring among the considered ones according to the limited feedback, the proposed adaptive compute-and-forward provides a better performance than that provided by the conventional compute-and-forward scheme which works over Gaussian or Eisenstein integers solely.
\end{abstract}


%------------------------------------------------------------------------
\section{Introduction}
Compute-and-forward \cite{nazer2011CF} is a novel paradigm of information forwarding which allows relay nodes to compute and then forward functions of messages by exploiting the structure induced by the channel. The main enabler of the scheme in \cite{nazer2011CF} is the use of lattice codes from Construction A over $\mbb{Z}$ the ring of integers. Later it was shown that compute-and-forward can also be performed over other rings of integers such as the Gaussian integers $\Zi$ and the Eisenstein integers $\Zw$ \cite{Feng10,Engin14,sun13} by constructing lattices via Construction A over the respective ring of integers.

%One of the important advantages of using lattices over $\Zi$ and $\Zw$ is that in effect it is possible to quantize the channel coefficients to elements in these rings and decode linear combinations of lattice codewords with coefficients chosen from the ring of integers used for constructing the lattice.

On one hand, since $\Zi$ and $\Zw$ are instances of imaginary quadratic integers, it seems natural to extend the compute-and-forward framework to general rings of imaginary quadratic integers. On the other hand, since the role of the underlying ring of integers can be effectively thought of as being a quantizer of the channel and $\Zw$ is already the best quantizer for $\mbb{C}$ where the channel coefficients belong, it seems unnecessary to pursue compute-and-forward over other rings of integers. In this paper, we first seek to better understand the role of rings of algebraic integers in constructing good lattices. We then use an example, namely compute-and-forward with limited feedback, to demonstrate the benefits of performing compute-and-forward over rings of imaginary quadratic integers other than $\Zi$ and $\Zw$.

One important difference between a general ring of imaginary quadratic integers and the Gaussian and Eisenstein integers is that the Gaussian integers and Eisenstein integers are not merely rings, they are principal ideal domains (PIDs). Hence, every ideal is generated by a singleton and one can equivalently work with numbers instead of ideals. The constructions of lattices over these two rings \cite{Feng10,Engin14,sun13} heavily rely on properties of PID. However, a general ring of imaginary quadratic integers is not a PID. Therefore, in order to fully exploit the potential of rings of imaginary quadratic integers, one has to work with ideals. In this paper, we generalize the famous Construction A to a general ring of imaginary quadratic integers (not necessary a PID). We show that such construction can produce lattices that are Poltyrev-good and MSE quantization-good. Based on this, we show that the rates expressed in \cite{nazer2011CF}, but with function coefficients from the adopted ring of imaginary integers can be achieved. %Based on this, the same achievable rate expression as that in \cite{nazer2011CF} with function coefficients chosen from the adopted ring of imaginary integers can be shown.

%Since $\Zw$ best quantizes $\mbb{C}$ among all imaginary quadratic integers, at first glance, it appears that there is no need to pursue lattices over rings other than $\Zw$. However, this is not the case when we have feedback.
Compute-and-forward with feedback was first studied in \cite{niesen11} in which the global channel knowledge is assumed at the transmitters. Using the theory of Diophantine approximation, Niesen and Whiting show that the traditional lattice-based scheme in \cite{nazer2011CF} is inefficient in the asymptotically high signal-to-noise ratio (SNR) regime. They then proposed a novel coding scheme which is a clever combination of compute-and-forward and real interference alignment \cite{motahari09}. Their scheme achieves the full degrees of freedom (DoF); but in order to see a gain, it requires an enormously high SNR. Another approach proposed in \cite{sakzad14} is to phase-precode the lattice-based scheme in \cite{nazer2011CF}. In this approach, one rotates the transmitted signal space according to the channel realization in such a way that the received signal space is close to a linear integer combination of the codebook. Hence, instead of the global channel knowledge, the phase-precoding approach only requires limited feedback. i.e., each transmitter only needs to know its optimal (or a reasonably good) phase for precoding.

In this paper, we propose a novel framework called adaptive compute-and-forward which makes use of the proposed lattices. The idea is to let the transmitters adaptively choose the best ring of imaginary quadratic integers to work with according to the channel coefficients. It is worth noting that this approach only requires the knowledge of which ring the transmitters should work with and hence, limited feedback suffices. We show that the proposed adaptive compute-and-forward can achieve increased computation rates as compared to the conventional compute-and-forward scheme in \cite{nazer2011CF}. Further, this can be used in conjunction with the phase precoding scheme in \cite{sakzad14}.

The idea of using different sets of algebraic integers for compute-and-forward was first proposed independently in \cite{Vazquez-Castro14, huangITW14}. In \cite{Vazquez-Castro14}, Vazquez-Castro uses finite constellations carved from some rings of imaginary quadratic integers which also form Euclidean domains (hence PIDs) for compute-and-forward. In \cite{huangITW14}, instead of being confined in Euclidean domains or PIDs, we go beyond PIDs and construct lattices over rings of imaginary quadratic integers for compute-and-forward. However, their goodness have not been shown and the idea of adaptively choosing the rings of integers was only vaguely mentioned in \cite{huangITW14}. This paper contributes to the literature by proving the optimality of the proposed lattices, and hence, deriving the achievable information rates with lattices over imaginary quadratic integers. The result of this paper also hints at why establishing converses or capacity results for compute-and-forward problems is difficult as they must allow for the possibility of \textit{many} rings of integers, not just Gaussian or Eisenstein integers.

The main contributions of this paper are also listed as follows:
\begin{itemize}
  \item While most of the work in the literature using large dimensional lattices (\cite{erez04,erez05,wilson10,nam10,ling14,zamir_book} for examples) focus mostly on constructions over PIDs ($\mbb{Z}$, $\Zi$, and $\Zw$ in particular) and that using algebraic integers for lattices focus mainly on small dimensional lattices (see \cite{joseph96,oggier04,oggier07} for example), the present work is one of the first in considering constructions of lattices/lattice codes over a general Dedekind domain. i.e., we combine the so far most popular constructions for the small and large dimensional lattices to construct asymptotically good lattices which inherit the algebraic structures of the underlying small dimensional lattices. We must emphasize that there is an independent work \cite{KositOggier15} which considers lattice codes via Construction A over number fields for block fading channel. This construction of lattices is quite general and subsumes our construction as a special case. However, due to different applications at hand, they largely focus on constructions over rings of integers of totally real number fields and do not prove the optimality of lattices thus constructed.
  \item To the best of our knowledge, this work presents the first results about the optimality of lattices constructed over non-PID domains. This paper advances our knowledge of lattice coding theory towards large dimensional random lattices generated from small dimensional ones which are full of structure and can serve as a stepping stone for further research. Recently, there have been some researches that identify the benefits of using lattice codes constructed over rings of algebraic integers other than imaginary quadratic integers (see \cite{KositOggier15} and \cite{campello16} for example). The proof techniques used in this paper may be extendable to prove the optimality for those lattices.
  \item This paper provides an application of the proposed lattices to compute-and-forward and proposes adaptive compute-and-forward. Simulation results show benefits of the proposed scheme which exploits the structure of proposed lattices and allows computation of functions with respect to the underlying ring of integers other than $\mbb{Z}$, $\Zi$, and $\Zw$. As mentioned above, Vazquez-Castro also proposed codes carved from lattices constructed over rings other than $\mbb{Z}$, $\Zi$, and $\Zw$ for compute-and-forward in \cite{Vazquez-Castro14}. However, the rings considered therein are limited to those belonging to Euclidean domains, which are again PID. More importantly, no optimality has been shown in \cite{Vazquez-Castro14}.
  %\item Although not discussed in details, the proposed approach can be easily generalized to constructions such as Construction D \cite{BarnesSloane83,conway1999sphere,forney2000}, Construction $\pi_A$, and Construction $\pi_D$ \cite{huang15} for generating multilevel lattices over algebraic integers.
\end{itemize}

\subsection{Notations}
Throughout the paper, $\mbb{R}$ and $\mbb{C}$ represent the set of real numbers and complex numbers, respectively. We use $j\defeq \sqrt{-1}$ to denote the imaginary unit. For a complex number $x=a+jb\in \mbb{C}$ where $a,b\in \mbb{R}$, $\bar{x}\defeq a-jb$ denotes its complex conjugate. We use $\Pp(E)$ to denote the probability of the event $E$. Vectors are written in boldface and random variables are written in Sans Serif font. We use $\times$ to denote the Cartesian product and use $\oplus$ and $\odot$ to denote the addition and multiplication operations, respectively, over a finite field where the field size can be understood from the context if it is not specified. Also, we do not distinguish the multiplication operation over the complex field and finite fields as it can be understood from the context.
%For a set $\mc{H}$ of vectors, $\mc{H}^*$ denotes the collection of non-zero elements in $\mc{H}$. i.e., $\mc{H}\setminus\mathbf{0}$.

\subsection{Organization}
The rest of the paper is organized as follows. In Section~\ref{sec:problem}, we discuss the compute-and-forward relay network which is a building block of large networks and is usually used to demonstrate the usefulness of the compute-and-forward technique. Construction A lattices over imaginary quadratic integers is proposed in Section~\ref{sec:lattices_Zd} followed by the proposed adaptive compute-and-forward scheme in Section~\ref{sec:acf}. Some concluding remarks are then given in Section~\ref{sec:conclude}. Since the proposed construction and the proposed adaptive compute-and-forward heavily rely on knowledge about lattices, algebra, and algebraic number theory, a brief review about these topics is given in Appendix~\ref{apx:prelim}. Also, we defer the proofs of the optimality of the proposed lattices to Appendix~\ref{apx:proof}.

\section{Problem Statement}\label{sec:problem}
\begin{figure}
    \centering
    \includegraphics[width=3.2in]{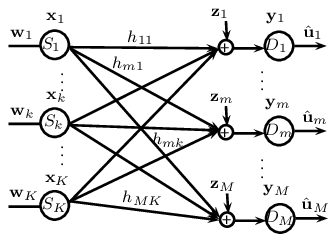}
    \caption{A compute-and-forward relay network where $S_1,\ldots,S_K$ are source nodes and $D_1,\ldots,D_M$ are destination nodes.}
    \label{fig:CF_model}
\end{figure}

The network considered in this paper is the compute-and-forward relay network studied by Nazer and Gastpar in \cite{nazer2011CF}. The network consists of $K$ source nodes and $M$ destination nodes as shown in Fig~\ref{fig:CF_model}. Each source node has a message $w_k\in\{1,2,\ldots,W\}$, $k\in\{1,\ldots,K\}$ which can alternatively be expressed by a length-$N'$ vector over some finite field, i.e., $\mathbf{w}_k\in\mbb{F}_p^{N'}$ with $W=p^{N'}$. This message is fed into an encoder $\mathcal{E}^N_k$ whose output is a length-$N$ codeword $\mathbf{x}_k\in\mbb{C}^N$. The codeword is subject to the average power constraint $P$ per complex dimension given by
\begin{equation}
    \frac{1}{N}\mbb{E}\| \mathbf{x}_k\|^2 \leq P.
\end{equation}
%\begin{equation}
%    \frac{1}{N}\| \mathbf{x}_k\|^2 = \frac{1}{N}\sum_{n=1}^N |x_k[n]|^2 \leq P\quad\text{per complex dimension.}
%\end{equation}

%The signal observed at destination $m$ is given by
%\begin{equation}
%    y_m[n] = \sum_{k=1}^K h_{mk}x_k[n] + z_m[n],
%\end{equation}
%where $h_{mk}\in\mbb{C}$ is the channel coefficient between the source node $k$ and destination $m$, and $z_m[n]\sim \mathcal{CN}(0,1)$. One can also define the channel model for using the channel $N$ times as
The received signal at destination $m$ is given by
\begin{equation}\label{eqn:y_m}
    \mathbf{y}_m = \sum_{k=1}^K h_{mk}\mathbf{x}_k + \mathbf{z}_m,
\end{equation}
where $h_{mk}\in\mbb{C}$ is the channel coefficient between the source node $k$ and destination node $m$, and $\mathbf{z}_m\sim \mathcal{CN}(0,\mathbf{I})$. One can think of this model as simply a layer in a large network; therefore, these destination nodes are merely intermediate relay nodes only interested in forwarding signals. Thus, instead of individual messages, each destination node is only interested in recovering a function of messages which will be forwarded to the next layer. In the Nazer and Gastpar's setting, functions are chosen to be linear combination of messages\footnote{In general, the functions are not limited to linear combinations of messages and some other functions have been considered in \cite{niesen11} \cite{Brett13} \cite{huang14isit} for instance.} given by
\begin{equation}
    \mathbf{u}_m = b_{m1}\odot \mathbf{w}_1\oplus\ldots\oplus b_{mK}\odot \mathbf{w}_K,
\end{equation}
where $b_{m1},\ldots,b_{mK}$ are elements in the same field $\mbb{F}_p$ with the elements in $\mathbf{w}_k$ and the operations are elementwise. Upon observing $\mathbf{y}_m$, the destination node $m$ forms $\hat{\mathbf{u}}_m = \mathcal{G}^N_m(\mathbf{y}_m)$ an estimate of $\mathbf{u}_m$.

\begin{define}[Computation codes]
    For a given equation coefficient vector $\mathbf{b}_m\defeq [b_{m1},\ldots,b_{mK}]^T$, a $(N,N')$ computation code consists of a sequence of encoding/decoding functions $(\mathcal{E}^N_1, \ldots, \mathcal{E}^N_K)/(\mathcal{G}^N_1, \ldots, \mathcal{G}^N_M)$ described above and an error probability given by
    \begin{equation}
        P_e^{(N)} \triangleq \Pp\left(\bigcup_{m=1}^M\left\{\hat{\mathbf{u}}_m \neq \mathbf{u}_m\right\}\right).
    \end{equation}
\end{define}

\begin{define}[Computation rate for function $\mathbf{b}_m$ at relay $m$]\label{def:com_rate_am}
    For a given channel vector $\mathbf{h}_m \triangleq [h_{m1},\ldots,h_{mK}]^T$ and equation coefficient vector $\mathbf{b}_m$, a computation rate $R(\mathbf{h}_m,\mathbf{b}_m,P)$ is achievable at relay $m$ if for any $\varepsilon>0$ there is an $(N,N')$ computation code such that
    \begin{equation}
        N'\geq NR(\mathbf{h}_m,\mathbf{b}_m,P)/\log(p) \text{~and~} P_e^{(N)}\leq \varepsilon.
    \end{equation}
    Note that the first condition is equivalent to saying that $W\geq 2^{N R(\mathbf{h}_m,\mathbf{b}_m,P)}$.
\end{define}
In this paper, we consider the symmetric case where all the encoders are of the same rate. Thus, for a given $\mathbf{H}\triangleq [\mathbf{h}_1,\ldots,\mathbf{h}_M]$ and $\mathbf{B}\triangleq [\mathbf{b}_1,\ldots,\mathbf{b}_M]$, the achievable computation rate is $R(\mathbf{H},\mathbf{B},P)\defeq \displaystyle{\min_{m}} R(\mathbf{h}_m,\mathbf{b}_m,P)$. Moreover, suppose there is a final destination collecting all the functions, it would be able to recover all the messages if $\mathbf{B}$ is full rank. Hence, one can also define the computation rate of the network as follows.
\begin{define}[Computation rate of the network]
    The achievable computation rate of the network is defined as
    \begin{equation}
        R(\mathbf{H},P) \triangleq \underset{\mathbf{B}:\mathbf{B}\text{~full rank}}{\max} R(\mathbf{H},\mathbf{B},P).
    \end{equation}
\end{define}

\begin{remark}
    The above definitions only consider the symmetric case in the sense that all the transmitters have the same power constraint $P$ and all the encoders have the same rate. For constructions over $\mbb{Z}$ lattices, some asymmetric cases have been discussed. For example, \cite{nazer2011CF} extends the compute-and-forward paradigm to case where transmitters have asymmetric rates. This extension is enabled by constructing a sequence of nested fine lattices and allowing each transmitter to choose a different fine lattice in this sequence. In \cite{ntranos13}, Ntranos \textit{et al.} further generalize the compute-and-forward paradigm to the scenario where transmitters may have unequal power constraints. This is done by allowing the transmitters to have different coarse and fine lattices. It is worth noting that the adaptive compute-and-forward scheme proposed in this paper can also be extended similarly to the asymmetric rates and asymmetric power constraints case. In this paper, we only present the symmetric result for the sake of conciseness.
\end{remark}

\subsection{Open-Loop Compute-and-Forward}
In what follows, we first consider the open-loop setting where channel state information is only available at the receivers. In \cite{nazer2011CF}, Nazer and Gastpar propose a novel paradigm called compute-and-forward where each source node implements the same nested lattice code over $\mbb{Z}$ of Erez and Zamir \cite{erez04} and encode real and imaginary parts separately. This allows each relay to decode the received signal to linear combination of the transmitted lattice points with coefficients being integers and results in the following computation rate at a relay.
\begin{theorem}[Nazer-Gastpar]\label{thm:comp_rate_Rm}
    At the $m$th relay, given $\mathbf{h}_m\in\mbb{C}^K$ and $\mathbf{a}_m\in\Zi^K$, the following computation rate is achievable\footnote{Note that here $\mathbf{a}_m\in\Zi^K$ but in Definition~\ref{def:com_rate_am}, computation rate is defined for $\mathbf{b}_m\in\mbb{F}_p^K$. This is not an issue by letting $p$ the field size tend to $\infty$, which is exactly what is required by the coding scheme in \cite{nazer2011CF}.}
    \begin{align}\label{eqn:open_CF_rate}
        R(\mathbf{h}_m,\mathbf{a}_m,P)=\log^{+}\left(\left(\|\mathbf{a}_m\|^{2}-
        \frac{P|\mathbf{h}_m^{H}\mathbf{a}_m|^2}{1+P\|\mathbf{h}_m\|^2}\right)^{-1}\right).
    \end{align}
\end{theorem}
Each relay can adaptively choose the coefficients $\mathbf{a}_m$ according to the channel vector $\mathbf{h}_m$ such that the above computation rate is maximized. Let $\mathbf{A}=[\mathbf{a}_1,\ldots,\mathbf{a}_M]$ and $\mathbf{B}=[\mathbf{b}_1,\ldots,\mathbf{b}_M]$ be its finite field representative. As long as the coefficient matrix $\mathbf{B}$ is full rank, the central destination, which has collected all the computed functions, is able to solve the linear equations and get the individual messages.

In \cite{nazer2011CF}, the real and the imaginary parts are separately considered.  In what follows, we provide a high-level description of the scheme for the real part only but the other part works identically. Each source node adopts a same nested lattice code constructed over $\mbb{Z}$. Since lattices are closed under integer linear combinations, the $m$th destination can now decode the codeword corresponding to an integer linear combination of codewords which were sent. Note that the channel output is a noisy version of a linear combination of codewords; hence, extra noise will be introduced when we try to enforce real linear combinations into integer linear combinations. This extra noise, called self-interference, can be equivalently represented as quantization error of quantizing a version of $\mathbf{h}_m$ with respect to the quantizer $\mbb{Z}^K$. After decoding $\sum_{k=1}^K a_{mk}\mathbf{x}_k$, the decoder then maps it back to the finite field and obtains $\hat{\mathbf{u}}_m$.

The proof of the above result heavily relies on two key points. The first one is the existence of good lattices from this construction. Perhaps more importantly, as recognized in \cite{Feng10}, the second one is that the mapping between $\mbb{Z}$ and the finite field is a ring homomorphism so that integer combinations of lattice points will be corresponding to linear combinations over finite field. This ring homomorphism then allows one to map back and forth between the finite field and real filed without ruining the structure. In \cite{Engin14}, Tunali $\textit{et al.}$ considered the real and the imaginary parts jointly and generalized the compute-and-forward paradigm to the Eisenstein lattices which are constructed from Construction A over $\Zw$. With this approach, the self-interference becomes quantization error of quantizing a version of $\mathbf{h}_m$ (now in $\mbb{C}^K$) with respect to the quantizer $\Zw^K$. This has resulted in an increased achievable rate on average as $\Zw$ approximates $\mbb{C}$ better than $\Zi$.

\section{Lattices over Imaginary Quadratic Integers}\label{sec:lattices_Zd}
Since both the Gaussian integers and Eisenstein integers are rings of integers of some number fields, it is natural to consider rings of integers of other number fields. In what follows, we particularly pick those rings of integers of imaginary quadratic fields. The reasons that we pick such rings are twofold. First of all, the channel coefficients we are trying to quantize lie in $\mbb{C}$, which is an extension field of $\mbb{R}$ with degree 2. Hence, it is natural to first investigate extensions with degree 2. Secondly, quadratic fields have been extensively studied and many properties have been discovered. This makes the generalization a lot easier. The discussions starting from this point heavily use knowledge of lattices, algebra, and algebraic number theory. For background knowledge on these topics, the reader is referred to Appendix~\ref{apx:prelim}.

Consider an imaginary quadratic field $\mbb{K}=\Qd$ with $d<0$ and its ring of integer $\Ok=\mbb{Z}[\xi]$ where
\begin{equation}\label{eqn:zbasis}
    \xi = \left\{\begin{array}{ll}
    \sqrt{d},                                           & d\equiv 2,3\mod 4, \\
    \frac{1+\sqrt{d}}{2},                                    & d\equiv 1\mod 4.\\
    \end{array} \right.
\end{equation}
We now discuss the Construction A lattices over $\Ok$ and some of the properties of such lattices. Note that since not every $\Ok$ forms a PID, one may have to work with ideals now instead of working with generators of ideals as we have done in PID cases. Let $\mfk{p}$ be a prime ideal that lies above $p$. The norm of the ideal $\mfk{p}$ denoted by $N(\mfk{p})$ is equal to $p^f$ where $f\in\{1,2\}$ is the inertial degree. One important property of $\Ok$ is that every prime ideal in $\Ok$ is also maximal. Hence,  $\Ok/\mfk{p}\cong \mbb{F}_{p^f}$ and we denote by $\mc{M}:\mbb{F}_{p^f} \rightarrow \Ok/\mfk{p}$ a ring isomorphism induced by this quotient ring.

{\bf \underline{Construction A}} \cite{LeechSloane71} \cite{conway1999sphere} Let $n$, $N$ be integers such that $n\leq N$ and let $\mathbf{G}$ be a generator matrix of an $(N,n)$ linear code over $\mbb{F}_{p^f}$. Construction A over $\Ok$ consists of the following steps:
\begin{enumerate}
    \item Define the discrete codebook $C=\{\mathbf{x}=\mathbf{G}\odot\mathbf{y}:\mathbf{y}\in\mbb{F}_{p^f}^n\}$ where all operations are over $\mbb{F}_{p^f}$.
    \item Construct $\Lambda^*\defeq \mc{M}(C)$ where $\mc{M}:\mbb{F}_{p^f}\rightarrow \Ok/\mfk{p}$ is a ring isomorphism.
    \item Tile $\Lambda^*$ to the entire $\mbb{C}^N$ to form $\Lambda\defeq \Lambda^* + \mfk{p}^N$.
\end{enumerate}

\begin{theorem}
    $\Lambda$ is a lattice over $\mbb{C}^N$. Moreover, a complex vector $\boldsymbol\lambda$ belongs to $\Lambda$ if and only if $\sigma(\boldsymbol\lambda)\in C$ where $\sigma\defeq \mc{M}^{-1}\circ\hspace{-3pt}\mod \mfk{p}^N$ is a ring homomorphism.
\end{theorem}
\begin{IEEEproof}
    Since $\mc{M}$ is a ring isomorphism, $\mc{M}(\mathbf{0})=\mathbf{0}$. Moreover, $\mathbf{0}\in\mfk{p}^N$. Thus, $\mathbf{0}\in\Lambda$. Let
    \begin{align}
        \boldsymbol\lambda_1 &= \mc{M}(\mathbf{c}_1) + \mathbf{p}_1, \\
        \boldsymbol\lambda_2 &= \mc{M}(\mathbf{c}_2) + \mathbf{p}_2,
    \end{align}
    where $\mathbf{c}_1,\mathbf{c}_1\in C$ and $\mathbf{p}_1,\mathbf{p}_2\in\mfk{p}^N$. We have
    \begin{align}
        \boldsymbol\lambda_1 + \boldsymbol\lambda_2 &= \mc{M}(\mathbf{c}_1) + \mc{M}(\mathbf{c}_2) + \mathbf{p}_1 +\mathbf{p}_2 \nonumber \\
        &\overset{(a)}{=} \mc{M}(\mathbf{c}_1\oplus \mathbf{c}_2) + \mathbf{p}+ \mathbf{p}_1 +\mathbf{p}_2 \nonumber \\
        &= \mc{M}(\mathbf{c}_3) + \mathbf{p}_3,
    \end{align}
    where $\mathbf{p},\mathbf{p}_3\in\mfk{p}^N$, $\mathbf{c}_3=\mathbf{c}_1\oplus\mathbf{c}_2\in C$ and (a) is due to the fact that $\mc{M}$ is a ring isomorphism. Moreover, choosing $\mathbf{c}_2$ to be the inverse of $\mathbf{c}_1$ and choosing $\mathbf{p}_2 = - \mathbf{p}_1 - \mathbf{p}$ makes $\boldsymbol\lambda_2$ the additive inverse of $\boldsymbol\lambda_1$. Therefore, $\Lambda$ is a lattice.

    To see that $\boldsymbol\lambda$ is a lattice point if and only if $\sigma(\boldsymbol\lambda)\in C$, we note that
    \begin{align}
        \hphantom{(\Leftrightarrow)}&~~\boldsymbol\lambda\in\Lambda \nonumber \\
        (\Leftrightarrow)&~~\boldsymbol\lambda = \mc{M}(\mathbf{c}) + \mathbf{p} \nonumber \\
        (\Leftrightarrow)&~~\boldsymbol\lambda \hspace{-3pt}\mod \mfk{p}^N = \mc{M}(\mathbf{c}) \nonumber \\
        (\Leftrightarrow)&~~\mc{M}^{-1}\left(\boldsymbol\lambda \hspace{-3pt}\mod \mfk{p}^N\right) = \mathbf{c}.
    \end{align}
\end{IEEEproof}

\begin{theorem}
    For any $d<0$ square free integer, consider $\mbb{K}=\mbb{Q}(\sqrt{d})$, there exists a sequence of lattices from Construction A over $\Ok$ that is Poltyrev-good and good for quantization.
\end{theorem}
\begin{IEEEproof}
    \textit{(Sketch. Please see Appendix~\ref{apx:proof} for details)} For showing Poltyrev-goodness, we tailor the Minkowski-Hlawka theorem specifically for $\Ok$ and then follow the steps of Loeliger in \cite{loeliger97} to show that with high probability, the random Construction A ensemble over $\Ok$ would produce Poltyrev-good lattices. For showing the MSE quantization-goodness, we modify the proof by Ordentlich and Erez \cite{ordentlich_erez_simple} where we first construct a sequence of prime ideals whose norms tend to $\infty$ for each $\Ok$ and show that randomly picking elements in $\mathbf{G}$ induces uniform distribution over $(\Ok/\mfk{p})^N$. One can then follow the steps in \cite{ordentlich_erez_simple} to show the MSE quantization-goodness.
\end{IEEEproof}

%Another famous construction of lattices is the so called Construction D \cite{BarnesSloane83,conway1999sphere,forney2000} which is a multilevel lattice construction. This construction can also be adapted to imaginary quadratic integers as follows.
%{\bf \underline{Construction D}} {\red Let $n^l$, $N$ be integers such that $n^l\leq N$ and let $\mathbf{G}_l$ be a generator matrix of an $(N,n^l)$ linear code over $\mbb{F}_{p_l^{f_l}}$. Construction $\pi_A$ over $\Ok$ consists of the following steps:
%\begin{enumerate}
%    \item Define the discrete codebook $C^l=\{\mathbf{x}=\mathbf{G}_l\odot\mathbf{y}:\mathbf{y}\in(\mbb{F}_{p_l^{f_l}})^n\}$ where all operations are over $\mbb{F}_{p_l^{f_l}}$.
%    \item Construct $\Lambda^*\defeq \mc{M}(C^1,\ldots,C^L)$ where $\mc{M}:\times_{l=1}^L\mbb{F}_{p_l^{f_l}}\rightarrow \Ok/\mfk{I}$ is a ring isomorphism.
%    \item Tile $\Lambda^*$ to the entire $\mbb{C}^N$ to form $\Lambda\defeq \Lambda^* + \mfk{I}^N$.
%\end{enumerate}}
%This multilevel lattice construction is expected to be able to produce multilevel lattices that are Poltyrev-good under multistage decoding.

Furthermore, similar to \cite{huang15}, one can even build multilevel lattices over $\Ok$ by Construction $\pi_A$ as follows. Let $\mfk{I}$ be a ideal whose prime ideal factorization is given by $\mfk{I}=\Pi_{l=1}^L \mfk{p}_l$ with $\mfk{p}_l$s relatively prime. From the Chinese remainder theorem, we have
\begin{align}
    \Ok/\mfk{I} &\cong \Ok/\Pi_{l=1}^L \mfk{p}_l \nonumber \\
    &\cong \Ok/\mfk{p}_1\times\ldots\times\Ok/\mfk{p}_L \nonumber \\
    &\cong \mbb{F}_{p_1^{f_1}}\times\ldots\times\mbb{F}_{p_L^{f_L}},
\end{align}
where $f_l$ is the inertial degree of $\mfk{p}_l$ in $\Ok$. We are ready to state Construction $\pi_A$ over $\Ok$.

{\bf \underline{Construction $\pi_A$}} Let $n^l$, $N$ be integers such that $n^l\leq N$ and let $\mathbf{G}_l$ be a generator matrix of an $(N,n^l)$ linear code over $\mbb{F}_{p_l^{f_l}}$. Construction $\pi_A$ over $\Ok$ consists of the following steps:
\begin{enumerate}
    \item Define the discrete codebook $C^l=\{\mathbf{x}=\mathbf{G}_l\odot\mathbf{y}:\mathbf{y}\in(\mbb{F}_{p_l^{f_l}})^n\}$ where all operations are over $\mbb{F}_{p_l^{f_l}}$.
    \item Construct $\Lambda^*\defeq \mc{M}(C^1,\ldots,C^L)$ where $\mc{M}:\times_{l=1}^L\mbb{F}_{p_l^{f_l}}\rightarrow \Ok/\mfk{I}$ is a ring isomorphism.
    \item Tile $\Lambda^*$ to the entire $\mbb{C}^N$ to form $\Lambda\defeq \Lambda^* + \mfk{I}^N$.
\end{enumerate}
Similar to Construction A, one can show that Construction $\pi_A$ over $\Ok$ always produces a lattice. One may also show that Construction $\pi_A$ over $\Ok$ can produce a sequence of lattices that are Poltyrev-good with high probability; however, the focus of this paper is on achievable computation rates rather than complexity so we do not pursue this. The interested reader is referred to \cite{huang15}.

\section{Proposed Adaptive Compute-and-Forward}\label{sec:acf}
In this section, we consider the scenario where there is \textit{limited feedback}. We propose the adaptive compute-and-forward scheme where we use the lattices proposed in Section~\ref{sec:lattices_Zd} to construct nested lattice codes and show that for each ring of imaginary quadratic integers, one can achieve the same rate expression \eqref{eqn:open_CF_rate} with elements in $\mathbf{a}_m$ chosen from that ring of integers. Depending on how many bits of feedback is available, the proposed adaptive compute-and-forward preselects a set of $\Ok$ and the idea is then simply to work with the ring of integers in this set that would result in the maximal computation rate. This approach can be used in conjunction with the phase-precoded compute-and-forward to further improve the performance.

%Here, we propose a novel coding strategy for the closed-loop setting. The idea is very simple; one should choose the best ring of integers to work with. We present the coding scheme and the achievable computation rate in the following.
For a given $\Ok$, we adapt the construction in \cite{ordentlich_erez_simple} to construct nested lattice codes over $\Ok$. Let $\mfk{p}$ be a prime ideal in $\Ok$ with $N(\mfk{p})=p\in (2N^3,2 \zeta N^3)$ a prime where $\zeta$ is a constant which guarantees the existence of such primes (see discussion in Appendix~\ref{apx:proof} for more details about $\zeta$). Let $\mc{M}:\mbb{F}_{p}\rightarrow \Ok/\mfk{p}$ be a ring isomorphism. It is worth noting that the scheme and the result for the other class of primes (those stay inert) can be obtained in a similar way with slight modification of parameters.

Let $(C_f,C_c)$ be a pair of nested linear code such that $C_c\subseteq C_f$ as follows,
\begin{align}
    C_c &=\left\{ \mathbf{G}_c\odot \mathbf{v} |\mathbf{v}\in\mbb{F}_{p}^{m_c}\right\}, \\
    C_f &=\left\{ \mathbf{G}_f\odot \mathbf{v} |\mathbf{v}\in\mbb{F}_{p}^{m_f}\right\},
\end{align}
where $\mathbf{G}_c$ is a $N\times m_c$ matrix and $\mathbf{G}_f = [\mathbf{G}_c~\tilde{\mathbf{G}}]$ with $\tilde{\mathbf{G}}$ being a $N\times (m_f-m_c)$ matrix. A pair of (scaled) nested lattice codes can be constructed by the construction described in Section~\ref{sec:lattices_Zd} as
\begin{align}
    \Lambda_c &= \gamma \mc{M}(C_c) + \gamma \mfk{p}, \\
    \Lambda_f &= \gamma \mc{M}(C_f) + \gamma \mfk{p},
\end{align}
where $\gamma \defeq \sqrt{4NP|d|^{-\frac{1}{2}}p^{-1}}$ is for power constraint. We then use $\Lambda_f\cap\mc{V}(\Lambda_c)$ as our nested lattice code whose design rate is given by
\begin{equation}
    R_{\text{design}} = \frac{m_f-m_c}{N}\log(p).
\end{equation}
Note that the difference between any two neighboring possible design rates is
\begin{align}
    \frac{1}{N}\log(p)&<\frac{1}{N}\log(2\zeta N^3) \nonumber \\
    &\leq \frac{\log(2\zeta)+3\log(N)}{N},
\end{align}
which can be made arbitrarily small as $N$ increases. Hence, the design rate can be set to approach any target rate.

Each source node adopts a same nested lattice code $\Lambda_f\cap\mc{V}(\Lambda_c)$ obtained by the above construction.  Specifically, for a message $\mathbf{w}_k\in\mbb{F}_p^{m_f-m_c}$, the overall encoding function acts as follows. We first pad $m_c$ zeros in the front to get $\mathbf{v}_k=q(\mathbf{w})=[\mathbf{0}_{m_c}^T,\mathbf{w}_k^T]^T$ and form the corresponding codeword $\mathbf{c}_{k}\in C_f$. This codeword is then mapped to a fine lattice codeword $\mathbf{t}_k=\gamma\mc{M}(\mathbf{c}_k)\hspace{-3pt}\mod\Lambda_c$. The transmitted signal at the source node $k$ is then given by
\begin{equation}
    \mathbf{x}_k = (\mathbf{t}_k -\mathbf{d}_k) \mod \Lambda_c,
\end{equation}
where $\mathbf{d}_k$ is a random dither.

According to the channel parameters, the destination node $m$ computes a linear combination of transmitted signals with coefficients $\mathbf{a}_m=[a_{m1},\ldots,a_{mK}]^T$ being elements in $\Ok$ and maps this function back to the finite field via the ring homomorphism corresponding to the inverse of $\mc{M}$.

\begin{figure}
    \centering
    \includegraphics[width=7in]{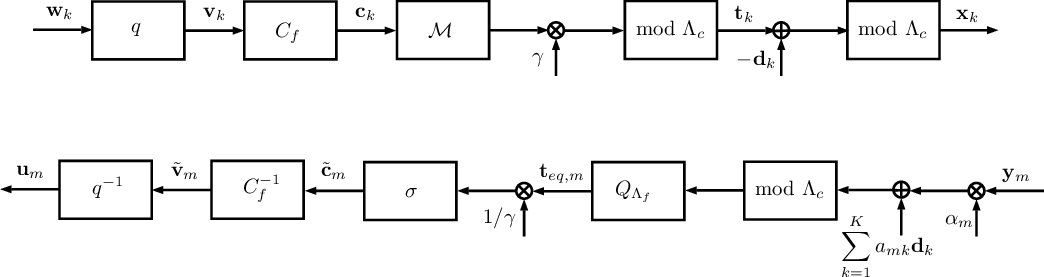}
    \caption{The encoding and decoding procedure of the proposed adaptive compute-and-forward scheme.}
    \label{fig:enc_dec}
\end{figure}

Specifically, the $m$-th receiver first forms
\begin{align}\label{eqn:NG_y}
    \mathbf{y}'_m &= \left(\alpha_m\mathbf{y}_m + \sum_{k=1}^K a_{mk}\mathbf{d}_k\right) \mod \Lambda_c \nonumber \\
    &= \left(\sum_{k=1}^K \alpha_mh_{mk}\mathbf{x}_k + a_{mk}\mathbf{d}_k+ \alpha_m\mathbf{z}_m  \right) \mod \Lambda_c \nonumber \\
    &= \left(\sum_{k=1}^K a_{mk}\mathbf{t}_{k} + \left(\alpha_m\mathbf{z}_m + \sum_{k=1}^K (\alpha_m h_{mk}-a_{mk})\mathbf{x}_k\right)  \right) \mod \Lambda_c \nonumber \\
    &= (\mathbf{t}_{eq,m} + \mathbf{z}_{eq,m}) \mod \Lambda_c,
\end{align}
where
\begin{equation}\label{eqn:NG_t_eq}
    \mathbf{t}_{eq,m}= \sum_{k=1}^K a_{mk}\mathbf{t}_{k} \mod \Lambda_c,
\end{equation}
is again a lattice codeword in $\Lambda_f\cap\mc{V}(\Lambda_c)$ since $\Lambda_f$ and $\Lambda_c$ are constructed over $\Ok$ and
\begin{equation}\label{eqn:NG_z_eq}
    \mathbf{z}_{eq,m} = \left(\alpha_m\mathbf{z}_m + \sum_{k=1}^K (\alpha_m h_{mk}-a_{mk})\mathbf{x}_k\right)\mod \Lambda_c,
\end{equation}
is the effective noise, also called the self-interference. It then performs lattice decoding to decode $\mathbf{y}'_m$ to the nearest element in $\Lambda_f$ to form $\hat{\mathbf{t}}_{eq,m}$. This estimated function is then mapped back to the finite field via the canonical ring homomorphism $\sigma \defeq \mc{M}^{-1}\circ\hspace{-3pt}\mod \mfk{p}$ to get $\hat{\mathbf{u}}_m=\sigma(\hat{\mathbf{t}}_{eq,m}/\gamma)$. Note that if $\hat{\mathbf{t}}_{eq,m}=\mathbf{t}_{eq,m}$, then
\begin{align}\label{eqn:u_NG}
    &\sigma\left(\frac{1}{\gamma}\left(\sum_{k=1}^K a_{mk}\mathbf{t}_{k} \mod \Lambda_c\right)\right) = \sigma\left(\frac{1}{\gamma}\left(\sum_{k=1}^K a_{mk}\mathbf{t}_{k} + \blambda_c\right)\right) \nonumber \\
    &= \sigma\left(\sum_{k=1}^K a_{mk}(\mc{M}(\mathbf{c}_k) + \mathbf{p}_k) + \mc{M}(\mathbf{c}_c) +\mathbf{p}_c \right) \nonumber \\
    &\overset{(a)}{=} \left(\bigoplus_{k=1}^K \sigma(a_{mk})\odot[\sigma (\mc{M}(\mathbf{c}_k))\oplus \sigma(\mathbf{p}_k)]\right) \oplus \sigma(\mc{M}(\mathbf{c}_c))\oplus\sigma(\mathbf{p}_c)  \nonumber \\
    &= \left(\bigoplus_{k=1}^K \sigma(a_{mk})\odot\sigma \left(\mc{M}(\mathbf{c}_{k})\right)\right) \oplus\sigma(\mc{M}(\mathbf{c}_c)) \nonumber \\
    &\overset{(b)}{=} \left(\bigoplus_{k=1}^K b_{mk} \odot \mathbf{c}_{k}\right) \oplus \mathbf{c}_c \nonumber \\
    &\overset{(c)}{=} \tilde{\mathbf{c}}_m\in C_f,
\end{align}
where $\mathbf{p}_k,\mathbf{p}_c\in\mfk{p}$, $\mathbf{c}_c\in C_c$, (a) is due to the fact that $\sigma$ is a ring homomorphism, (b) follows from $b_{mk} = \sigma(a_{mk})\in\mbb{F}_{p}$, and (c) is due to the linearity of $C_f$ and the fact that $\mathbf{c}_c\in C_c\subseteq C_f$. Since there is a one-to-one mapping between the input and output of the linear code $C_f$, one can then invert the linear code $C_f$ to get $\tilde{\mathbf{v}}_m$ corresponding to $\tilde{\mathbf{c}}_m$. Moreover, since $\mathbf{c}_c\in C_c$, it only affects the first $m_c$ positions of $\tilde{\mathbf{v}}_m$. Hence, $\tilde{\mathbf{v}}_m$ and $\bigoplus_{k=1}^K b_{m,k}\odot\mathbf{v}_k$ lie inside the same coset. We then remove the first $m_c$ positions via $q^{-1}(.)$ to get the corresponding function of messages as
\begin{equation}
    \mathbf{u}_m =q^{-1}(\tilde{\mathbf{v}}_m)= b_{m1} \odot \mathbf{w}_1\oplus\ldots\oplus b_{mK} \odot \mathbf{w}_K.
\end{equation}
The overall encoding and decoding procedure is summarized in Fig.~\ref{fig:enc_dec}

Let us write $\mathbf{h}_m=[h_{m1},\ldots,h_{mK}]^T$ and $\mathbf{a}_m=[a_{m1},\ldots,a_{mK}]^T$. Using the goodness results in Section~\ref{sec:lattices_Zd} and choosing $\alpha_m$ to be the MMSE estimator given by
\begin{equation}
    \alpha_m = \frac{P\mathbf{h}_m^H\mathbf{a}_m}{1+\|\mathbf{h}_m\|^2},
\end{equation}
one can follow the steps in \cite{nazer2011CF} \cite{ordentlich_erez_simple} to show that the following computation rate is achievable\footnote{We slightly abuse notation by allowing elements in $\mathbf{a}_m$ from $\Ok$ while in Definition~\ref{def:com_rate_am} it should be $\mathbf{b}_m$ from $\mbb{F}_p$. One can get around with this issue by noting that $\mathbf{b}_m=\sigma(\mathbf{a}_m)$ and letting $p\rightarrow\infty$.}
\begin{equation}\label{eqn:com_rate_m}
        R_{\Ok}(\mathbf{h}_m,\mathbf{a}_m,P) = \log^+\left(\left(\|\mathbf{a}_m\|^2-\frac{P|\mathbf{h}_m^*\mathbf{a}_m|^2}{1+P\|\mathbf{h}_m\|^2}\right)^{-1}\right).
    \end{equation}
where $\log^+(.)\defeq \max\{0,\log(.)\}$. i.e., as $N$ increases, we can operate at a design rate $R_{\text{design}}>R_{\Ok}(\mathbf{h}_m,\mathbf{a}_m,P)-\varepsilon$ for any $\varepsilon>0$ with vanishing $p_e^{(N)}$.

Note that here the subscript $\Ok$ is used to emphasize that this is obtained by working over a particular ring of imaginary quadratic integers $\Ok$. Suppose we have a limited feedback of $\nu$ bits, then we can pre-select $\mc{A}$, a set of $2^{\nu}$ rings of imaginary integers. To obtain the highest computation rate for the proposed framework, one then solves the following optimization problem to decide which $\Ok\in\mc{A}$ to work with.
\begin{equation}\label{eqn:R_opt_ok}
    R(\mathbf{H},P) = \max_{\Ok\in\mc{A}} \max_{\sigma\left(\mathbf{A}\right)\text{~invertible}} R_{\Ok}(\mathbf{H},\mathbf{A},P),
\end{equation}
where $R_{\Ok}(\mathbf{H},\mathbf{A},P) = \min_m R_{\Ok}(\mathbf{h}_m,\mathbf{a}_m,P)$.

\subsection{Numerical Results}
\begin{figure}
    \centering
    \includegraphics[width=3.5in]{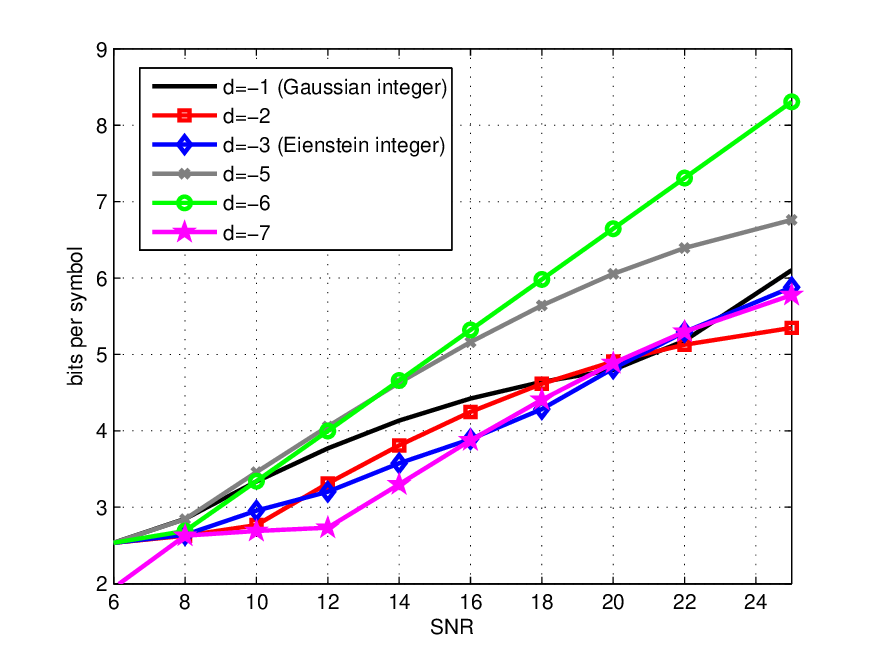}
    \caption{Comparison of computation rates for different $\Ok$ for the 2 by 2 case where $h_{11}=h_{22} = 1$ and $h_{12}= h_{21}=j 2.449$.}
    \label{fig:fix_ch_d_6}
\end{figure}

We now provide two numerical results to demonstrate the benefits of using the proposed lattices. For both the results, we consider the case where there are 2 source nodes and 2 destination nodes. In Fig.~\ref{fig:fix_ch_d_6}, we consider fixed channel coefficients $h_{11}=h_{22} = 1$ and $h_{12}= h_{21}=j 2.449$ and show the achievable computation rates obtained by using lattices over $\Ok$ of $\mbb{Q}(\sqrt{d})$ for $d\in\{-1,-2,-3,-5,-6,-7\}$. One can see from Fig.~\ref{fig:fix_ch_d_6} that although $\Zw$ the ring of Eisenstein integers best approximates $\mbb{C}$ among all imaginary quadratic integers, for specific channel coefficients, it is possible that there are other rings of integers which have elements closer to those channel coefficients than $\Zw$ does. In this example, $\mbb{Z}[\sqrt{-6}]$ has elements closer to the specific channel coefficients than other rings considered in this simulation and hence provide the best computation rate among them in the high SNR regime.

\begin{figure}
    \centering
    \includegraphics[width=3.5in]{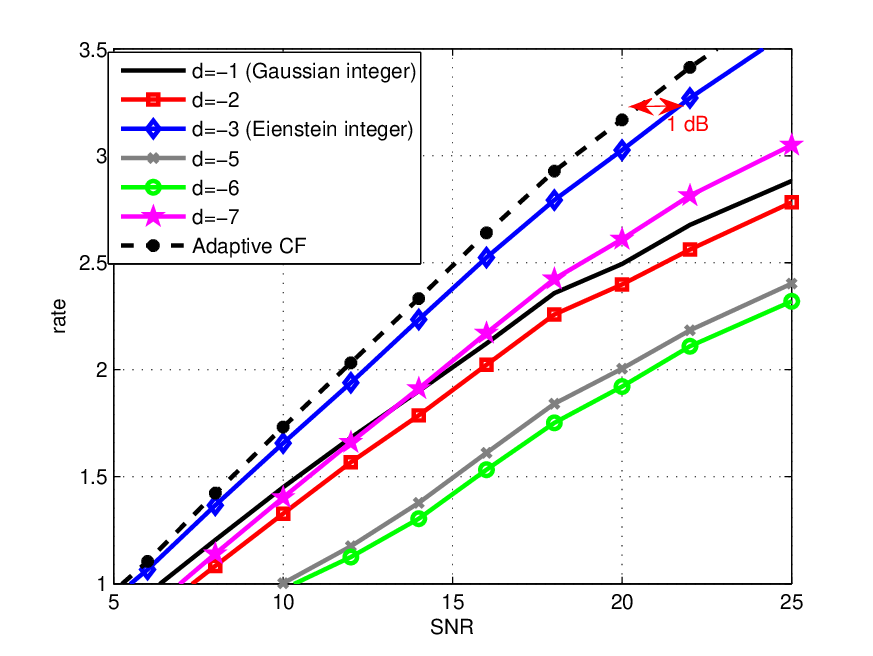}
    \caption{Comparison of average computation rates for different $\Ok$. The average is taken over 10000 pairs of channel realizations drawn from Rayleigh distribution.}
    \label{fig:avg_ch}
\end{figure}

In Fig.~\ref{fig:avg_ch}, we provide average achievable computation rates by using lattices over $\Ok$ of $\mbb{Q}(\sqrt{d})$ for $d\in\{-1,-2,-3,-5,-6,-7\}$. In this figure, each channel coefficient is randomly drawn from circularly symmetric Gaussian distribution with variance 1. i.e., its norm has Rayleigh distribution. We average over 10000 realizations and show that on average alternating between these 6 rings provides better performance than that provided by working over any of them individually.

\subsection{Discussion}
We now discuss some extensions and issues for the proposed scheme. As mentioned above, one can incorporate the idea of phase-precoding into the proposed adaptive compute-and-forward framework to further improve the overall performance. Specifically, according to the channel realization, for each ring of imaginary quadratic integers, one can first examine the computation rate achieved by phase-precoding the proposed nested lattice codes constructed over that ring of integers. One can then choose the ring that leads to the highest computation rate to work with. Note that phase-precoded compute-and-forward with integers can be thought of as using rotated integers to approximate the original channel coefficients \cite{sakzad14}. But there are algebraic integers which cannot be expressed as rotated integers; hence in general, allowing working over other rings of algebraic integers will result in an increased computation rate. Another way to exploit this is to first determine the ring of integers and then rotate the integers by phase-precoding. Either way will result in further reduction of the self-interference.

%Another possibility is to consider lattices over number fields with degree larger than 2. Since the channel coefficients that we try to approximate lie in $\mbb{C}$ the field extension of $\mbb{R}$ with degree 2, it seems to suggest that there is no need to go beyond $\Ok$ with degree 2. However, for the separation-based compute-and-forward framework where one uses a linear code in conjunction with a constellation carved from a lattice, the shaping gain is completely determined by the constellation; hence, it makes perfect sense to study $\Ok$ with a higher degree. To this end, constellations carved from rings of cyclotomic integers for the separation-based compute-and-forward are currently under investigation.

A potential weakness of the proposed framework is the complexity issue. This comes from two different aspects. Firstly, the optimization problem in \eqref{eqn:R_opt_ok} is in general very difficult to solve. Fortunately, good approximation algorithms have been proposed for some rings of integers \cite{Feng10,Engin14}. These algorithms are based on LLL lattice basis reduction algorithm \cite{LLL}. The extension of the LLL algorithms to many other rings of integers can be found in \cite{Napias} for example. Moreover, simulation results shown above suggest that one does not have to consider too many $\Ok$ for getting improved performance. Secondly, the modulo operation with respect to an ideal may cause increased complexity. Luckily, there are algorithms available which have polynomial running time. For example, \cite[Algorithm 1.4.12]{Cohen00} will produce a unique canonical coset representative very efficiently. Furthermore, for those $\Ok$ which also form Euclidean domains (there are exactly five of them corresponding to $d\in\{-1,-2,-3,-7,-11\}$), one can further reduce the complexity by taking advantage of Euclidean functions as reported in \cite{Vazquez-Castro14}.

\section{Conclusion}\label{sec:conclude}
In this paper, we have moved beyond PIDs and generalized Construction A of lattices to general rings of algebraic integers of imaginary quadratic fields. We have then shown that such construction can produce good lattices in the sense of Poltyrev and MSE quantization. When used for compute-and-forward, these lattices have allowed us to reliably compute linear combinations of codewords with coefficients being elements in the underlying ring which the lattices are constructed over. A novel scheme named adaptive compute-and-forward has been proposed where the transmitter first chooses a ring of algebraic integers depending on the channel coefficients and then uses a lattice code over the chosen ring of algebraic integers. This allows us to obtain higher computation rates than using a fixed lattice code. Moreover, one can phase-precode the proposed adaptive compute-and-forward scheme to further improve the performance.

%\section*{Acknowledge}
%The authors would like to thank Prof. Bobak Nazer for his question that motivated this work. The authors are thankful to Prof. Uri Erez for the discussion about showing the goodness for the proposed lattices during his visit at TAMU.

\appendices
\section{Preliminaries}\label{apx:prelim}
In this appendix, we review the literature on lattices, algebra, and algebraic number theory that will be the foundation of this work. All the Lemmas are provided without proofs for the sake of brevity; however, their proofs can be found in standard textbooks. For details, please see for example \cite{zamir_book,erez04,erez05,Hungerford74,StewardTall,lang94}.

\subsection{Lattices}
An $N$-dimensional lattice $\Lambda$ is a discrete subgroup of $\mathbb{R}^N$ which satisfies the following properties: $\mathbf{0}\in\Lambda$, $\forall \boldsymbol\lambda\in\Lambda$, we have $-\boldsymbol\lambda\in\Lambda$, and  $\forall \boldsymbol\lambda_1, \boldsymbol\lambda_2\in \Lambda$, we have $\boldsymbol\lambda_1 + \boldsymbol\lambda_2 \in \Lambda$. Some important operations and notions for lattices are defined as follows. For a $\mathbf{x}\in\mathbb{R}^N$, the nearest neighbor quantizer associated with $\Lambda$ is denoted as
\begin{equation}
    Q_{\Lambda}(\mathbf{x})=\boldsymbol\lambda\in\Lambda;~\|\mathbf{x}-\boldsymbol\lambda\|\leq\|\mathbf{x}-\boldsymbol\lambda'\|~\forall\boldsymbol\lambda'\in\Lambda,
\end{equation}
where $\| .\|$ represents the $L_2$-norm operation and the ties are broken arbitrarily. After defining lattice quantization, we can define the fundamental Voronoi region $\mathcal{V}_{\Lambda}$ as $\mathcal{V}_{\Lambda}=\{ \mathbf{x}: Q_{\Lambda}(\mathbf{x})=\mathbf{0} \}.$ The $\hspace{-3pt}\mod \Lambda$ operation simply provides quantization error with respect to $\Lambda$ and is represented as
\begin{equation}
    \mathbf{x}\hspace{-3pt}\mod \Lambda = \mathbf{x}-Q_{\Lambda}(\mathbf{x}).
\end{equation}

The second moment of a lattice is defined as
\begin{equation}
    \sigma^2(\Lambda) = \frac{1}{\text{Vol}(\mc{V}_{\Lambda})}\frac{1}{N}\int_{\mathcal{V}_{\Lambda}}\| \mathbf{x} \|^2 \mathrm{d}\mathbf{x},
\end{equation}
where $\text{Vol}(\mc{V}_{\Lambda})$ is the volume of $\mathcal{V}_{\Lambda}$ and the normalized second moment of the lattice is then defined as
\begin{equation}
    G(\Lambda) = \frac{\sigma^2(\Lambda)}{\text{Vol}(\mc{V}_{\Lambda})^{2/N}}.
\end{equation}
The normalized second moment is a dimensionless quantity and is invariant to scaling. A lower bounded on $G(\Lambda)$ can be obtained by the normalized second moment of a ball which asymptotically approaches $\frac{1}{2\pi e}$ in the limit as $N\rightarrow \infty$.

We are ready to define the two goodness properties considered in this work, namely the goodness for MSE quantization and goodness for channel coding (Poltyrev-goodness).
\begin{define}[Goodness for MSE Quantization]
    A sequence of lattices is asymptotically good for MSE quantization if
    \begin{equation}
        \underset{N\rightarrow\infty}{\lim} G(\Lambda) = \frac{1}{2\pi e}.
    \end{equation}
\end{define}
Consider the communication channel $\mathbf{y}=\mathbf{x}+\mathbf{z}$ where each element in $\mathbf{z}$ is i.i.d.$\sim \mathcal{N}(0,\eta^2)$ and there is no power constraint on $\mathbf{x} \in \Lambda$. The Poltyrev-goodness is defined as follows.
\begin{define}[Poltyrev-Goodness]
    A sequence of lattices is asymptotically Poltyrev-good if whenever
    \begin{equation}\label{eqn:poltyre_good}
        \eta^2<\frac{\text{Vol}(\mc{V}_{\Lambda})}{2\pi e},
    \end{equation}
    the error probability of decoding $\mathbf{x}$ from $\mathbf{y}$ can be made arbitrarily small.
\end{define}

\subsection{Algebra}
Let $\mc{R}$ be a commutative ring. Let $a, b\neq 0 \in\mc{R}$ but $ab = 0$, then $a$ and $b$ are \textit{zero divisors}. If $ab = ba = 1$, then we say $a$ is a \textit{unit}. Two elements $a, b\in\mc{R}$ are associates if $a$ can be written as the multiplication of a unit and $b$. A non-unit element $\tau\in\mc{R}$ is a prime if whenever $\tau$ divides $ab$ for some $a, b \in \mc{R}$, either $\tau$ divides $a$ or $\tau$ divides $b$. An \textit{integral domain} is a commutative ring with identity and no zero divisors. An additive subgroup $\mfk{I}$ of $\mc{R}$ satisfying $ar\in\mfk{I}$ for $a\in\mfk{I}$ and $r\in\mc{R}$ is called an \textit{ideal} of $\mc{R}$. An ideal $\mfk{I}$ of $\mc{R}$ is proper if $\mfk{I}\neq\mc{R}$. An ideal generated by a singleton is called a \textit{principal ideal}. A \textit{principal ideal domain} (PID) is an integral domain in which every ideal is principal. Famous and important examples of PID include $\mbb{Z}$, $\Zi$ and $\Zw$. Let $a, b\in\mc{R}$ and $\mfk{I}$ be an ideal of $\mc{R}$; then $a$ is congruent to $b$ \textit{modulo} $\mfk{I}$ if $a-b\in\mfk{I}$. The quotient ring $\mc{R}/\mfk{I}$ of $\mc{R}$ by $\mfk{I}$ is the ring with addition and multiplication defined as
\begin{align}
    (a+\mfk{I})+(b+\mfk{I}) &= (a+b)+\mfk{I}, \text{~and} \\
    (a+\mfk{I})\cdot(b+\mfk{I}) &= (a\cdot b)+\mfk{I}.
\end{align}
A proper ideal $\mfk{p}$ of $\mc{R}$ is said to be a \textit{prime ideal} if for $a, b\in\mc{R}$ and $ab\in\mfk{p}$, then either $a\in\mfk{p}$ or $b\in\mfk{p}$. A proper ideal $\mfk{I}$ of $\mc{R}$ is said to be a \textit{maximal ideal} if $\mfk{I}$ is not contained in any strictly larger proper ideal. It should be noted that every maximal ideal is also a prime ideal but the reverse may not be true. Let $\mc{R}_1, \mc{R}_2, \ldots, \mc{R}_L$ be a family of rings, the direct product of these rings, denoted by $\mc{R}_1\times \mc{R}_2\times \ldots \times\mc{R}_L$, is the direct product of the additive abelian groups $\mc{R}_l$ equipped with multiplication defined by the \textit{componentwise} multiplication.

Let $\mfk{I}_1$ and $\mfk{I}_2$ be two ideals of $\mc{R}$, we shall now define some operations of ideals. The sum of two ideals is the ideal defined as
\begin{equation}
    \mfk{I}_1+\mfk{I}_2 \defeq \left\{a+b: a\in\mfk{I}_1, b\in\mfk{I}_2\right\}.
\end{equation}
Two ideals are \textit{relatively prime} if $\mc{R} = \mfk{I}_1+\mfk{I}_2$. The product of two ideals is the ideal defined as
\begin{equation}
    \mfk{I}_1\mfk{I}_2 \defeq \left\{ \sum_{i=1}^{n} a_i b_i: a_i\in\mfk{I}_1, b_i\in\mfk{I}_2, n\in\mbb{N} \right\}.
\end{equation}
In general, we have $\mfk{I}_1\mfk{I}_2 \subseteq \mfk{I}_1 \cap \mfk{I}_2$; but if they are relatively prime, then $\mfk{I}_1\mfk{I}_2 = \mfk{I}_1\cap\mfk{I}_2$. We say $\mfk{I}_1$ divides $\mfk{I}_2$ or $\mfk{I}_1|\mfk{I}_2$ if there is an ideal $\mfk{I}_3$ such that $\mfk{I}_2=\mfk{I}_1\mfk{I_3}$ (this is equivalent to $\mfk{I}_2\subseteq \mfk{I}_1$).

%Two ideals are \textit{relatively prime} if
%\begin{equation}
%    \mc{R} = \mfk{I}_1+\mfk{I}_2 \triangleq \{a+b: a\in\mfk{I}_1, b\in\mfk{I}_2\}.
%\end{equation}

Let $\mc{R}_1$ and $\mc{R}_2$ be rings. A function $\sigma:\mc{R}_1\rightarrow \mc{R}_2$ is a \textit{ring homomorphism} if
\begin{align}
    \sigma(a + b) &= \sigma(a) \oplus \sigma(b) ~\forall a,b\in\mc{R}_1 \text{~and} \\
    \sigma(a\cdot b) &= \sigma(a)\odot \sigma(b),~\forall a,b\in\mc{R}_1,
\end{align}
where $(+,\cdot)$ and $(\oplus,\odot)$ are operations in $\mc{R}_1$ and $\mc{R}_2$, respectively. A homomorphism is said to be \textit{monomorphism} if it is injective and \textit{isomorphism} if it is bijective. Let $\mc{R}$ be a commutative ring, and $\mfk{I}_1,\ldots,\mfk{I}_n$ be ideals in $\mc{R}$. Then, from the Chinese Remainder Theorem, we have
\begin{equation}
    \mc{R}/\cap_{i=1}^n\mfk{I}_i \cong \mc{R}/\mfk{I}_1\times\ldots\times\mc{R}/\mfk{I}_n.
\end{equation}
%A $\mc{R}$-module $\mc{N}$ over a ring $\mc{R}$ consists of an abelian group ($\mc{N},+$) and an operation $\mc{R}\times \mc{N}\rightarrow \mc{N}$ which satisfies the same axioms as those for vector spaces. Let $\mc{N}_1$ and $\mc{N}_2$ be $\mc{R}$-modules. A function $\varphi:\mc{N}_1\rightarrow \mc{N}_2$ is a \textit{$\mc{R}$-module homomorphism} if
%\begin{align}
%    \varphi(a + b) &= \varphi(a) + \varphi(b) ~\forall a,b\in\mc{N}_1 \text{~and} \\
%    \varphi(r\cdot a) &= r \cdot \varphi(b),~\forall r\in\mc{R}, b\in\mc{N}_1.
%\end{align}

%We now present some lemmas which serve as the foundation of the proposed constellations.
%\begin{lemma}\label{lma:PID}
%    If $\mc{R}$ is a PID, then every non-zero prime ideal is maximal.
%\end{lemma}

%\begin{lemma}\label{lma:MAX}
%    Let $\mfk{I}$ be an ideal in a ring $\mc{R}$ with identity $1_{\mc{R}}\neq 0$. If $\mfk{I}$ is maximal and $\mc{R}$ is commutative, then the quotient ring $\mc{R}/\mfk{I}$ is isomorphic to a field.
%\end{lemma}

%\begin{lemma}[Chinese Remainder Theorem]\label{lma:CRT}
%    Let $\mc{R}$ be a commutative ring, and $\mfk{I}_1,\ldots,\mfk{I}_n$ be ideals in $\mc{R}$, such that they are relatively prime. Then,
%    \begin{equation}
%        \mc{R}/\cap_{i=1}^n\mfk{I}_i \cong \mc{R}/\mfk{I}_1\times\ldots\times\mc{R}/\mfk{I}_n.
%    \end{equation}
%\end{lemma}

\subsection{Algebraic Numbers and Algebraic Integers}
Now, we provide some background knowledge on algebraic number theory. The materials are mostly from \cite{StewardTall} \cite{lang94} and proofs and algorithms can be found therein.

\begin{define}[Algebraic Numbers and Algebraic Number Fields]
    An algebraic number is a root of some polynomial with coefficients in $\mbb{Z}$. The set of all algebraic numbers is a subfield $\mbb{A}$ of $\mbb{C}$. We define a number field to be a subfield $\mbb{K}$ of $\mbb{A}$ (hence a subfield of $\mbb{C}$) such that the degree $[\mbb{K}:\mbb{Q}]$ is finite.
\end{define}
Theorem 2.2 in \cite{StewardTall} shows that any such $\mbb{K}$ is equal to $\mbb{Q}(\theta)$, the smallest subfield containing $\mbb{Q}$ and $\theta$, for some algebraic number $\theta$.

\begin{define}[Algebraic Integers]
    An algebraic integer is a complex number which is a root of some monic polynomial (whose leading coefficient is 1) with coefficients in $\mbb{Z}$. The set of all algebraic integers forms a subring $\mc{B}$ of $\mbb{C}$. For any number field $\mbb{K}$, we write $\mfk{O}_{\mbb{K}}=\mbb{K}\cap\mc{B}$ and call $\mfk{O}_{\mbb{K}}$ the ring of integers of $\mbb{K}$.
\end{define}
From Corollary 2.12 in \cite{StewardTall}, one has that if $\mbb{K}$ is a number field then $\mbb{K}=\mbb{Q}(\theta)$ for an algebraic integer $\theta$ which is called a primitive element for $\mbb{K}$ over $\mbb{Q}$. Also, in general, there will be several distinct $\mbb{Q}$-monomorphisms (i.e., it fixes $\mbb{Q}$) embedding $\mbb{K}$ into $\mbb{C}$. From Theorem 2.4 in \cite{StewardTall}, we know that for $\mbb{K}=\mbb{Q}(\theta)$ a number field of degree $n$ over $\mbb{Q}$, there are exactly $n$ distinct $\mbb{Q}$-monomorphism $\sigma_i:\mbb{K}\rightarrow\mbb{C}$ and such monomorphisms form a group $\text{Gal}(\mbb{K}/\mbb{Q})\defeq\{\sigma_1,\ldots,\sigma_n\}$ which is referred to as the Galois group. Moreover, for $\alpha\in\mbb{Q}(\theta)$, $\sigma_i(\alpha)$ for $i\in\{1,2,\ldots,n\}$ are the distinct zeros in $\mbb{C}$ of the minimal polynomial of $\alpha$ over $\mbb{Q}$. We call those $\sigma_i(\alpha)$ the \textit{conjugates} of $\alpha$ and define the \textit{norm} of $\alpha$ to be the product of conjugates as
\begin{equation}
    N_{\mbb{K}}(\alpha)=\prod_{i=1}^n\sigma_i(\alpha).
\end{equation}

Let $\{\alpha_1,\ldots,\alpha_n\}$ be a $\mbb{Q}$-basis for $\mbb{K}$. We define the \textit{discriminant} of $\{\alpha_1,\ldots,\alpha_n\}$ as
\begin{equation}
    \Delta[\alpha_1,\ldots,\alpha_n]\defeq \det\left(
                                                 \begin{array}{cccc}
                                                   \sigma_1(\alpha_1) & \sigma_1(\alpha_2) & \ldots & \sigma_1(\alpha_n) \\
                                                   \sigma_2(\alpha_1) & \sigma_2(\alpha_2) & \ldots & \sigma_2(\alpha_n) \\
                                                   \vdots & \vdots & \ddots & \vdots \\
                                                   \sigma_n(\alpha_1) & \sigma_n(\alpha_2) & \ldots & \sigma_n(\alpha_n) \\
                                                 \end{array}
                                               \right)^2.
\end{equation}
If $\{\alpha_1,\ldots,\alpha_n\}$ is a $\mbb{Z}$-basis for $\mfk{O}_\mbb{K}$, we define the discriminant of $\mbb{K}$ to be $\Delta_\mbb{K}\defeq \Delta[\alpha_1,\ldots,\alpha_n]$. Let $\mfk{I}$ be an ideal of $\mfk{O}_{\mbb{K}}$. The norm of $\mfk{I}$ is $N(\mfk{I})\defeq|\mfk{O}_{\mbb{K}}/\mfk{I}|$. Moreover, if $\{\beta_1,\ldots,\beta_n\}$ is a $\mbb{Z}$-basis for $\mfk{I}$, then $N(\mfk{I})=\sqrt{\frac{\Delta[\beta_1,\ldots,\beta_n]}{\Delta_{\mbb{K}}}}$. The norm is multiplicative, i.e., for two ideals $\mfk{I}_1$ and $\mfk{I}_2$ of $\Ok$, $N(\mfk{I}_1\mfk{I}_2)=N(\mfk{I}_1)N(\mfk{I}_2)$. It can be shown that if $N(\mfk{p})$ is a rational prime, then $\mfk{p}$ is a prime ideal.

In this paper, we are particularly interested in imaginary quadratic fields and their algebraic integers whose definitions can be found below.
\begin{define}[Quadratic Fields]
    A quadratic field is an algebraic number field $\mbb{K}$ of degree $[\mbb{K}:\mbb{Q}]=2$ over $\mbb{Q}$. Particularly, one may write $\mbb{K}=\mbb{Q}(\sqrt{d})$ where $d\in\mbb{Z}$ is square free. We say $\mbb{K}$ is an imaginary quadratic field if $d<0$.
\end{define}
Let $\mbb{K}=\mbb{Q}(\sqrt{d})$. One can show that $\mfk{O}_{\mbb{K}}=\mbb{Z}[\xi]$ where
\begin{equation}
    \xi = \left\{\begin{array}{ll}
    \sqrt{d},                                           & d\equiv 2,3\mod 4, \\
    \frac{1+\sqrt{d}}{2},                                    & d\equiv 1\mod 4.\\
    \end{array} \right.
\end{equation}
Also, $\Delta_\mbb{K}=4d$ if $d\equiv 2,3\mod 4$ and $\Delta_\mbb{K}=d$ if $d\equiv 1\mod 4$. It can be easily seen that when $d=-1$ we have the Gaussian integers and when $d=-3$ we have the Eisenstein integers.

\begin{example}
    Let us consider the case $d=-5$, i.e., $\mbb{K}=\mbb{Q}(\sqrt{-5})$. Let $\alpha=a + b \sqrt{-5}$ where $a,b\in\mbb{Z}$. Since the degree is 2, there are exactly two $\mbb{Q}$-monomorphisms. In order to have a $\mbb{Q}$-monomorphism, one must have $\sigma(\sqrt{-5})\cdot\sigma(\sqrt{-5}) = \sigma(-5) = -5$, which implies that $\sigma(\sqrt{-5})=\pm \sqrt{-5}$. Thus one has that
    \begin{align}
        \sigma_1(\alpha) &= a+b\sigma_1(\sqrt{-5}) = a+b\sqrt{-5}, \nonumber \\
        \sigma_2(\alpha) &= a+b\sigma_2(\sqrt{-5}) = a-b\sqrt{-5}.
    \end{align}
    Then the norm of $\alpha$ is $\sigma_1(\alpha)\cdot\sigma_2(\alpha)= a^2+5b^2$ which coincides with the Euclidean norm. Since $-5\equiv 3 \mod 4$, from \eqref{eqn:zbasis}, $\{1,\sqrt{-5}\}$ is a $\mbb{Z}$-basis for $\Ok$. One can calculate the discriminant as follows,
    \begin{equation}
        \Delta_{\mbb{K}}=\det\left(
                           \begin{array}{cc}
                             1 & \sqrt{-5} \\
                             1 & -\sqrt{-5} \\
                           \end{array}
                         \right)^2 = -20.
    \end{equation}
\end{example}

For any prime $p$, $p\mbb{Z}$ is a prime ideal in $\mbb{Z}$. Let $\mfk{p}$ be a prime ideal of $\Ok$. We say $\mfk{p}$ lies above $p$ if $\mfk{p}|p\mbb{Z}$. The ideal $p\Ok$ can be uniquely factorized into $p\Ok = \Pi_{l=1}^L \mfk{p}_l^{e_l}$ with $\mfk{p}_l$ distinct. We call $e_l$ the \textit{ramification index} of $\mfk{p}_l$ over $p$ and $f_l=[\Ok/\mfk{p}_l:\mbb{Z}/p\mbb{Z}]$ the \textit{inertial degree} of $\mfk{p}_l$ over $p$. Note that one must have $N(\mfk{p}_l)=p^{f_l}$. Also, the ramification indices and inertial degrees must satisfy $\sum_{l=1}^L e_lf_l=n$. If $e_l>1$ for some $l$, we say $p$ (or $p\Ok$ to be precise) ramifies in $\Ok$. If $L>1$, we say $p$ splits in $\Ok$. If $L=1$ and $e_1=1$ (i.e., $f_1=n$), we say $p$ remains inert in $\Ok$. The following Lemma allows one to efficiently categorize primes.

\begin{lemma}\label{lma:prime_category}
    Let $p$ be a rational prime. For a quadratic field $\mbb{K}$, one has
    \begin{itemize}
        \item if $\left(\frac{\Delta_{\mbb{K}}}{p}\right)=0$, then $p$ ramifies in $\Ok$,
        \item if $\left(\frac{\Delta_{\mbb{K}}}{p}\right)=1$, then $p$ splits in $\Ok$,
        \item if $\left(\frac{\Delta_{\mbb{K}}}{p}\right)=-1$, then $p$ remains inert in $\Ok$,
    \end{itemize}
    where $\left(\frac{\Delta_{\mbb{K}}}{p}\right)$ is the Kronecker symbol $\hspace{-3pt}\mod p$. Moreover, the Kronecker symbol $\hspace{-3pt}\mod p$ operation can be efficiently computed. (See for example \cite[Algorithm 1.4.10]{Cohen93}.)
% with running time $O(\log^2(\max\{\Delta_{\mbb{K}},p\}))$
\end{lemma}

It is well-known that not every ring of imaginary quadratic integers forms a PID. In fact, it was conjectured by Gauss and shown by Heegner and Stark that there are only 9 of them that are PIDs (corresponding to $d\in\{-1,-2,-3,-7,-11,-19,-43,-67,-163\}$). Therefore, it is crucial to have a systematic way to identify prime ideals. The following theorem provides a means of doing this.
\begin{lemma}
    Let $p$ be an odd rational prime. For a quadratic field $\mbb{K}=\mbb{Q}(\sqrt{d})$, one has
    \begin{itemize}
        \item if $p$ ramifies in $\Ok$, then $\mfk{p}=(p,\sqrt{d})$ is a prime ideal lying above $p$,
        \item if $p$ splits in $\Ok$, then $\mfk{p}=(p,a+\sqrt{d})$ is a prime ideal lying above $p$ for any $a$ such that $a^2\equiv d\mod p$.
    \end{itemize}
    Moreover, such $a$ can be efficiently found (See for example \cite[Algorithm 1.5.1]{Cohen93}.)
\end{lemma}
One important property of $\Ok$ is that every prime ideal is maximal. Therefore, we have that for every prime ideal $\mfk{p}$ in $\Ok$,
\begin{equation}
    \Ok/\mfk{p}\cong \mbb{F}_{p^f},
\end{equation}
where $f$ is the inertial degree described above.

\begin{example}
    Again consider $d=-5$, i.e., $\mbb{K}=\mbb{Q}(\sqrt{-5})$, and $p=23$. We have that $\left(\frac{\Delta_{\mbb{K}}}{p}\right)=1$; hence, $23\mbb{Z}$ splits into two prime ideals in $\Ok$. From the above theorem, one can check that $8^2\equiv -5\mod 23$; thus, $\Ok=\mfk{p}\bar{\mfk{p}}$ where $\mfk{p}=(23,8+\sqrt{-5})$. Also, we have
    \begin{equation}
        \Delta=\det\left(
                           \begin{array}{cc}
                             23 & 8+\sqrt{-5} \\
                             23 & 8-\sqrt{-5} \\
                           \end{array}
                         \right)^2 = -10580.
    \end{equation}
    Therefore, $N(\mfk{p})=\sqrt{-10580/-20}=23$. Moreover, $\Ok/\mfk{p}\cong \mbb{F}_{23}$. This coset decomposition and the corresponding ring isomorphism is shown in Fig.~\ref{fig:Z_5_homo}.

    \begin{figure}
    \centering
    \includegraphics[width=3.5in]{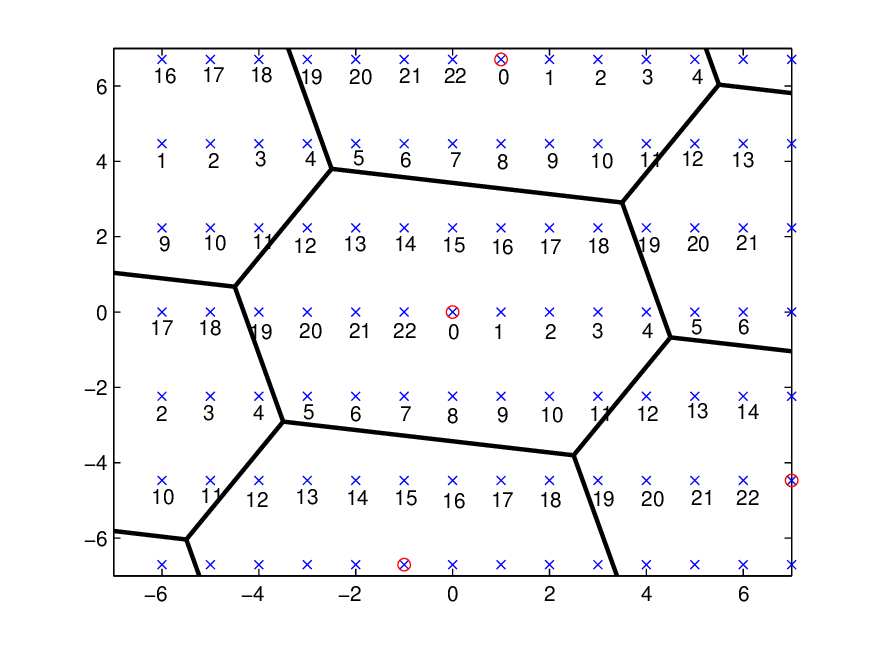}
    \caption{The coset decomposition and the corresponding ring isomorphism.}
    \label{fig:Z_5_homo}
    \end{figure}
\end{example}

\begin{lemma}[Dirichlet's prime theorem]\label{lma:dirichlet}
    For any two relatively prime integers $a$ and $d$, there are infinitely many rational primes of the form $p\equiv a\hspace{-3pt}\mod d$.
\end{lemma}
Note that for any odd prime $p$, $\left(\frac{\Delta_{\mbb{K}}}{p}\right)$ only depends on $p\hspace{-3pt}\mod 4\Delta_{\mbb{K}}$. Hence, Lemma~\ref{lma:prime_category} and Lemma~\ref{lma:dirichlet} together imply that for any quadratic field $\mbb{K}$, there exist infinitely many splitting primes and infinitely many inert primes as well. Moreover, for $\Ok$ of $\mbb{Q}(\sqrt{d})$ with $d\equiv 2,3\hspace{-3pt}\mod 4$, the behavior of its primes is regulated by
\begin{equation}
    \left(\frac{\Delta_{\mbb{K}}}{p}\right) = \left(\frac{4d}{p}\right) = \left(\frac{4}{p}\right)\left(\frac{d}{p}\right) = \left(\frac{d}{p}\right),
\end{equation}
where we have used the multiplicative property of Kronecker symbol and the fact that $\left(\frac{4}{p}\right)=1$. Therefore, for this class of rings, the behavior of the prime $p$ only depends on $p\hspace{-3pt}\mod 4d$. In what follows, we provide a weaker version of the Chebotarev's density theorem which further tells us how those primes distribute asymptotically.

\begin{lemma}[Chebotarev's density theorem]\label{lma:chebotarev}
    In a ring of algebraic integers of a quadratic field, asymptotically, the density of each category of primes is 1/2, i.e., asymptotically, half of the rational primes split and half of them remain inert.
\end{lemma}

\section{Proofs}\label{apx:proof}
In this appendix, we show that Construction A over $\Ok$ produces good lattices with high probability. Throughout the proof, we only consider $\Ok$, the ring of algebraic integers for $\mbb{Q}(\sqrt{d})$ with $d\equiv 2,3\hspace{-3pt}\mod 4$; the case of $d\equiv 1\hspace{-3pt}\mod 4$ can be proved similarly with a slight modification of parameters. Also, we focus on those primes which split completely in $\Ok$ for the sake of simplicity; however, the proof techniques can be applied to the proposed construction with primes that remain inert.

\subsection{Poltyrev-Goodness}
The proof closely follows the steps in \cite{loeliger97}. Let $p$ be a splitting prime in $\Ok$, i.e., $p\Ok$ splits into two prime ideals in $\Ok$, namely $p\Ok=\mfk{p}\bar{\mfk{p}}$. Therefore, we have $\Ok/\mfk{p}\cong \Ok/\bar{\mfk{p}}\cong \mbb{F}_p$. Let $\mc{C}$ be the collection of all $(N,n)$ linear codes $C$ over $\mbb{F}_p$. The set $\mc{C}$ is a balanced set and the basic averaging lemma \cite[Lemma 1]{loeliger97} applies. Thus, one has

\begin{equation}
    \frac{1}{|\mc{C}|}\sum_{C\in\mc{C}}\sum_{\mathbf{c}\in C\setminus \mathbf{0}} f(\mc{M}(\mathbf{c})) = \frac{p^n-1}{p^N-1}\sum_{\mathbf{s}\in(\mbb{F}_{p}^N)\setminus \mathbf{0}} f(\mc{M}(\mathbf{s})),
\end{equation}
for an arbitrary mapping $f:\mbb{C}^N\rightarrow \mbb{R}$. Since we can identify $\mbb{C}$ by $\mbb{R}^2$, the basic averaging lemma works for arbitrary mapping $f:\mbb{R}^{2N}\rightarrow \mbb{R}$ as well and we use $\mbb{C}^N$ and $\mbb{R}^{2N}$ interchangeably in the following.

\begin{theorem}[Modified Minkowski-Hlawka Theorem]
    Let $f:\mbb{R}^{2N}\rightarrow\mbb{R}$ be a Riemann integrable function of bounded support. Then, for any integer $0<n<N$, and any fixed $\text{Vol}(\mc{V}_{\Lambda})$, the approximation
    \begin{equation}
        \frac{1}{|\mc{C}|}\sum_{C\in\mc{C}}\sum_{\mathbf{v}\in\gamma\Lambda\setminus \mathbf{0}} f(\mathbf{v}) \approx \text{Vol}(\mc{V}_{\gamma\Lambda})^{-1}\int_{\mbb{R}^{2N}} f(\mathbf{v})d\mathbf{v},
    \end{equation}
    becomes exact in the limit $p\rightarrow\infty$, $\gamma^2(\frac{\sqrt{|\Delta_{\mbb{K}}|}}{2})\rightarrow 0$, $\text{Vol}(\mc{V}_{\gamma\Lambda})=\gamma^{2N} (\frac{\sqrt{|\Delta_{\mbb{K}}|}}{2})^N p^{N-n}$ fixed.
\end{theorem}
Before proceeding to the proof, we first note that due to Dirichlet's prime theorem (Lemma~\ref{lma:dirichlet}) and Chebotarev's density theorem (Lemma~\ref{lma:chebotarev}), there exist infinitely many splitting primes in every $\Ok$ so that one can safely let $p$ go to infinity.
\begin{IEEEproof}
    Recall that $\sigma\defeq \mc{M}^{-1}\circ\hspace{-3pt}\mod \mfk{p}^N$. Note that by the basic averaging lemma,
    \begin{align}
        &\frac{1}{|\mc{C}|}\sum_{C\in\mc{C}}\sum_{\mathbf{v}\in\gamma\Lambda\setminus \mathbf{0}} f(\mathbf{v}) = \frac{1}{|\mc{C}|}\sum_{C\in\mc{C}}\left[\sum_{\mathbf{v}\in(\Ok^N)\setminus \mathbf{0}:\sigma(\mathbf{v})=\mathbf{0}} f(\gamma\mathbf{v})\right. \nonumber \\
        &\hspace{4cm}\left.+ \sum_{\mathbf{v}\in(\Ok^N)\setminus \mathbf{0}:\sigma(\mathbf{v})\in C\setminus \mathbf{0}} f(\gamma \mathbf{v}) \right] \nonumber \\
        &= \sum_{\mathbf{v}\in(\Ok^N)\setminus \mathbf{0}:\sigma(\mathbf{v})=\mathbf{0}} f(\gamma \mathbf{v}) \nonumber \\
        &\hspace{3cm} + \frac{1}{|\mc{C}|}\sum_{C\in\mc{C}}\sum_{\mathbf{c}\in C\setminus \mathbf{0}}\left[\sum_{\mathbf{v}\in\Ok^N:\sigma(\mathbf{v})=\mathbf{c}} f(\gamma \mathbf{v}) \right] \nonumber \\
        &= \sum_{\mathbf{v}\in(\Ok^N)\setminus \mathbf{0}:\sigma(\mathbf{v})=\mathbf{0}} f(\gamma \mathbf{v}) \nonumber \\
        &\hspace{3cm} + \frac{p^n-1}{p^N-1} \sum_{\mathbf{c}\in \mbb{F}_p^N\setminus\mathbf{0}} \left[\sum_{\mathbf{v}\in\Ok^N:\sigma(\mathbf{v})=\mathbf{c}} f(\gamma \mathbf{v}) \right] \nonumber \\
        &= \sum_{\mathbf{v}\in(\Ok^N)\setminus \mathbf{0}:\sigma(\mathbf{v})=\mathbf{0}} f(\gamma \mathbf{v}) + \frac{p^n-1}{p^N-1} \sum_{\mathbf{v}\in\Ok^N:\sigma(\mathbf{v})\neq\mathbf{0}} f(\gamma \mathbf{v}) \nonumber \\
        &\overset{(a)}{\approx} p^{n-N} \gamma^{-2N} \left(\frac{\sqrt{|\Delta_{\mbb{K}}|}}{2}\right)^{-N}\nonumber \\
        &\hspace{3cm}\cdot\sum_{\mathbf{v}\in(\gamma\Ok)^N:\sigma(\mathbf{v})\neq\mathbf{0}} f(\mathbf{v}) \gamma^{2N}\left(\frac{\sqrt{|\Delta_{\mbb{K}}|}}{2}\right)^N \nonumber \\
        &\overset{(b)}{\approx} \text{Vol}(\mc{V}_{\gamma\Lambda})^{-1}\int_{\mbb{R}^{2N}} f(\mathbf{v})d\mathbf{v},
    \end{align}
    where (a) requires $\gamma p$ being large and $f$ having bounded support and (b) requires $\gamma^2(\frac{\sqrt{|\Delta_{\mbb{K}}|}}{2})$ to be small so that the Riemann sum approaches the Riemann integral.
\end{IEEEproof}
One can then follow the proof in \cite{loeliger97} to show that with high probability, the proposed construction produces lattices that are Poltyrev-good.

\subsection{MSE Quantization-Goodness}\label{apx:MSE_good}
The proof closely follows the steps in \cite{ordentlich_erez_simple}. We again consider splitting primes that can be decomposed into $p=\mfk{p}\bar{\mfk{p}}$ and use the proposed construction with the prime ideal $\mfk{p}$. Denote by $V_N$ the volume of an $N$-real dimensional ball with unit radius and let $\mc{B}(\mathbf{s},r)$ be a $2N$-real dimensional ball in (or equivalently $N$-complex dimensional ball) with radius $r$ centered at $\mathbf{s}$. We again prove the result for $p$ primes splitting completely in $\Ok$ only. i.e., $N(\mfk{p})=p$ for a prime ideal $\mfk{p}$ lying above $p$. Note that scaling would not change lattice structure; in the sequel, we equivalently consider the scaled version
\begin{equation}
    \Lambda = \gamma \mc{M}(C) + \gamma \mfk{p}^N,
\end{equation}
where $\gamma \defeq \sqrt{4NP|d|^{-\frac{1}{2}}p^{-1}}$ and we pick $2N^3\leq p\leq 2\zeta N^3$ where $\zeta$ is a constant to guarantee that we can find a prime $p$ splitting in $\Ok$. Note that as mentioned in Appendix~\ref{apx:prelim}, the behavior of primes in imaginary quadratic integers is regulated by the Kronecker symbol $\left(\frac{d}{p}\right)$ and depends only on $p\hspace{-3pt}\mod 4d$ for the class of rings considered here. i.e., we can find at least one $a$ such that every natural prime $p$ in the arithmetic progression $p\equiv a\hspace{-3pt}\mod D$ splits completely. The prime number theorem for arithmetic progressions \cite{moree93} states that for any $\zeta>1$, there must exist a sufficiently large $c$ such that for every $z\geq c$, there's at least one $p\in(z,\zeta z)$ in this arithmetic progression. However, it may not be an easy task to characterize how large $c$ has to be. Fortunately, the results in \cite[Table I]{moree93} provide small upper bound on $c$ with small $\zeta$ for many cases of $D\leq 840$, which is far more than enough for the application of adaptive compute-and-forward. %Note that since here we constrain $p$ to those splitting, one cannot use the Bertrand's postulate to guarantee that for any $N$, there always exists a $\xi \in [1/2,1)$ such that $\xi^2N^3$ is a prime square as done in \cite{ordentlich_erez_simple}. Although there are generalized versions of the Bertrand's postulate that may solve the problem for some $\Ok$ (see for example \cite{moree93} and reference therein), we do not purse this as a bound would be sufficient. %However, we can turn the table around and first decide the prime square $q$ and then find a suitable $N$. Since there exist infinitely many of inert primes for every $\Ok$, one can still let $q$, and hence $N$, go to infinity.
% $q \geq (2N)^3/4$

\begin{lemma}[Modified Lemma 1 in \cite{ordentlich_erez_simple}]\label{lma:lma1_or}
    For any $\mathbf{s}\in\mbb{R}^{2N}$ and $r>0$, the number of points of $\Ok^N$ inside $\mc{B}(\mathbf{s},r)$ can be bounded as
    \begin{align}
        &\left(\max\{r-\frac{\sqrt{2N|d|}}{2},0\}\right)^{2N} \cdot \frac{V_{2N}}{\left(\frac{\sqrt{|\Delta_{\mbb{K}}|}}{2}\right)^{N}}  \leq\left|\Ok^N\cap \mc{B}(\mathbf{s},r)\right| \nonumber \\
        &\hspace{2.3cm}\leq\left(r+\frac{\sqrt{2N|d|}}{2}\right)^{2N} \cdot \frac{V_{2N}}{\left(\frac{\sqrt{|\Delta_{\mbb{K}}|}}{2}\right)^{N}}.
    \end{align}
\end{lemma}
\begin{IEEEproof}
    Let $\mc{S}\defeq \Ok^N\cap\mc{B}(\mathbf{s},r) + \mc{V}_{\Ok^N}$ where $\mc{V}_{\Ok^N}$ is the fundamental Voronoi region of $\Ok^N$. One has
    \begin{align}
        \text{Vol}(\mc{S}) &= |\Ok^N\cap\mc{B}(\mathbf{s},r)|\cdot \left(\frac{\sqrt{|\Delta_{\mbb{K}}|}}{2}\right)^N.
    \end{align}
    Note that the largest distance between any two points lying within $\mc{V}_{\Ok^N}$ can be upper-bounded by $\sqrt{2N|d|}$. For any $\mathbf{x}\in\mc{B}(\mathbf{s},r-\sqrt{2N|d|}/2)$ that lies within $\mathbf{a}+\mc{V}_{\Ok^N}$ for some $\mathbf{a}\in\Ok^N$, we have
    \begin{equation}
        \|\mathbf{a}-\mathbf{x}\|\leq \frac{\sqrt{2N|d|}}{2}.
    \end{equation}
    Therefore, following the triangle inequality, we have $\|\mathbf{a}-\mathbf{s}\|\leq r$ and thus $\mathbf{x}\in\mc{S}$. This shows $\mathbf{x}\in\mc{B}(\mathbf{s},r-\sqrt{2N|d|}/2) \subseteq \mc{S}$. Moreover, we have
    \begin{equation}
        \mc{S}\subseteq \mc{B}(\mathbf{s},r) + \mc{V}_{\Ok^N} \subseteq \mc{B}(\mathbf{s},r) + \mc{B}(\mathbf{0},\sqrt{2N|d|}/2),
    \end{equation}
    which shows that $\mc{S}\subseteq \mc{B}(\mathbf{s},r+\sqrt{2N|d|}/2)$. Evaluating the volumes of these sets completes the proof.
\end{IEEEproof}
%\begin{IEEEproof}
%    Similar to \cite[Lemma 1]{ordentlich_erez_simple} and hence omitted.
%\end{IEEEproof}

For any $\mathbf{x}\in\mbb{C}^N$, define
\begin{align}
    d(\mathbf{x},\Lambda) &= \frac{1}{2N} \min_{\blambda\in\Lambda } \| \mathbf{x}-\blambda \|^2 \nonumber \\
    &= \frac{1}{2N} \min_{\mathbf{a}\in \mfk{p}^N,\mathbf{c}\in C} \| \mathbf{x}-\gamma \mc{M}(\mathbf{c})- \gamma \mathbf{a}\|^2 \nonumber \\
    &= \frac{1}{2N} \min_{\mathbf{c}\in C} \| (\mathbf{x}-\gamma \mc{M}(\mathbf{c}))^*\|^2,
\end{align}
where $y^*\defeq y\hspace{-3pt}\mod \gamma \mfk{p}^N$. Also, note that
\begin{align}
    d(\mathbf{x},\Lambda)&\leq \frac{\gamma^2 d_{\mfk{p}}^2}{4} \leq \frac{\gamma^2p\sqrt{|d|}}{2\pi},
\end{align}
where $d_{\mfk{p}}^2$ is the minimum squared Euclidean distance of elements in $\mfk{p}$ and the lower bound is due from a bound on $d_{\mfk{p}^2}$ \cite{peikert07,huang15LIC}.
%\begin{equation}
%    d(\mathbf{x},\Lambda)\leq \left\{\begin{array}{ll}
%    \frac{\gamma^2|d|}{4} & d\equiv 2,3\hspace{-3pt}\mod 4, \\
%    \frac{\gamma^2(1+|d|)}{16} & d\equiv 1\hspace{-3pt}\mod 4.
%    \end{array}\right.
%\end{equation}
Recall that for the case considered ($d\equiv 2,3\hspace{-3pt}\mod 4$), $\Delta_{\mbb{K}}=4d$. For any $\mathbf{w}\in\mbb{F}_p^n\setminus \mathbf{0}$, define the random vector $C(\mathbf{w})=\mathbf{G}\odot\mathbf{w}^T$ which is uniformly distributed over $\mbb{F}_p^N$. Thus, $\mc{M}(C(\mathbf{w}))$ is uniformly distributed over $(\Ok/\mfk{p})^N$.

For all $\mathbf{w}\in\mbb{F}_p^n\setminus \mathbf{0}$ and $\mathbf{x}\in\mbb{C}^N$, we have
\begin{align}\label{eqn:ep_bound}
    \varepsilon &\defeq \Pp \left( \frac{1}{2N}\|( \mathbf{x}-\gamma \mc{M}(C(\mathbf{w})) )^*\|^2\leq \frac{P}{2} \right) \nonumber \\
    &= p^{-N} \left| \gamma (\Ok/\mfk{p})^N \cap \mc{B}^*(\mathbf{x},\sqrt{NP}) \right| \nonumber \\
    &= p^{-N} \left| \gamma \Ok^N \cap \mc{B}(\mathbf{x},\sqrt{NP}) \right| \nonumber \\
    &\overset{(a)}{\geq} p^{-N}\left(\gamma^{-1}\sqrt{NP}-\frac{\sqrt{2N|d|}}{2}\right)^{2N} \cdot \frac{V_{2N}}{\left(\sqrt{|d|}\right)^{N}} \nonumber \\
    &= V_{2N}(\gamma^{-2}NP|d|^{-\frac{1}{2}}p^{-1})^N\left( 1-\frac{\gamma\sqrt{|d|}}{\sqrt{2P}} \right)^{2N} \nonumber \\
    &\overset{(b)}{=} V_{2N} 2^{-2N} \left( 1 - \frac{\sqrt{2N|d|^{1/2}}}{\sqrt{p}} \right)^{2N} \nonumber \\
    &\overset{(c)}{\geq} V_{2N} 2^{-2N} \left( 1 - \frac{|d|^{1/4}}{N} \right)^{2N} \nonumber \\
    %&> V_{2N} 2^{-2N} \left( 1 - \frac{\sqrt{|d|}}{N} \right)^{2N} \nonumber \\
    &\overset{(d)}{>} \frac{1}{(2N)^2}V_{2N}2^{-2N},
\end{align}
where (a) is from Lemma~\ref{lma:lma1_or}, (b) is due to the choice $\gamma = \sqrt{4NP|d|^{-\frac{1}{2}}}$, (c) is due to the choice $p\geq 2N^3$, and (d) is true for sufficiently large $N$. This can be verified by noting that $( 1 - |d|^{1/4}/N )^{2N}$ is a positive strictly increasing function for $N \geq |d|^{1/4}$ and will converge to $\exp(-2|d|^{1/4})>0$ while $1/(2N)^2$ is a positive strictly decreasing function and will converge to 0.

Let $W=p^n-1$ and label each of the $\mathbf{w}\in\mbb{F}_p^n\setminus \mathbf{0}$ by $i=1,\ldots,W$. Define the indicator random variable related to $\mathbf{x}\in\mbb{C}^N$ as
\begin{equation}
    \chi_i\defeq \left\{\begin{array}{ll}
    1, & \frac{1}{2N}\|(\mathbf{x}-\gamma p^{-1/2}\mc{M}(C_i))^*\|^2\leq P \\
    0, & \text{otherwise} \\
    \end{array}\right.
\end{equation}
One has that for any $\mathbf{x}\in\mbb{C}^N$
\begin{align}
    \Pp\left( (d(\mathbf{x},\Lambda))>P \right) &= \Pp\left(\sum_{i=1}^W \chi_i=0\right) \nonumber \\
    &\leq \Pp\left(\left|\sum_{i=1}^W \chi_i-\varepsilon\right|\geq \varepsilon\right) \nonumber \\
    &\overset{(a)}{\leq} \frac{\var(\frac{1}{W}\sum_{i=1}^W \chi_i)}{\varepsilon^2} \nonumber \\
    &= \frac{1}{W^2\varepsilon^2}\sum_{i=1}^W\sum_{l=1}^W \textrm{Cov}(\chi_i,\chi_l)\nonumber \\
    &\overset{(b)}{\leq} \frac{p}{W\varepsilon} \overset{(c)}{<} \zeta(2N)^5 p^{-n} 2^{2N}V_{2N}^{-1},
\end{align}
where (a) follows from the Chebyshev's inequality, (b) is due to the fact that $C(\mathbf{w}_1)$ and $C(\mathbf{w}_2)$ are statistically independent unless $\mathbf{w}_1=a\cdot \mathbf{w}_2$ for $a\in\mbb{F}_p$, and (c) is by plugging in \eqref{eqn:ep_bound} and the fact that $p\leq 2\zeta N^3$.

One can then show that for any distribution on $\msf{X}$, we have
\begin{align}
    &\mbb{E}_{\msf{X},\Lambda}(d(\msf{X},\Lambda)) \nonumber \\%&=\int_{d(\msf{X},\Lambda)<\frac{P}{2}} d(\msf{X},\Lambda)\Pp(d(\msf{X},\Lambda)) + \int_{d(\msf{X},\Lambda)>\frac{P}{2}} d(\msf{X},\Lambda)\Pp(d(\msf{X},\Lambda)) \nonumber \\
    &\hspace{0cm}\leq \frac{P}{2}\left( 1+\delta(2N)^6 2^{-N\left(\frac{n}{N}\log(p)-\log\left(\frac{4}{V_{2N}^{2/2N}}\right)\right)} \right),
\end{align}
where $\delta = \zeta/\pi$ is a constant. This in turn implies that
\begin{equation}
    \lim_{N\rightarrow \infty} \mbb{E}_{\Lambda}\left(\sigma^2(\Lambda)\right) \leq \frac{P}{2}\quad\text{per real dimension},
\end{equation}
if one chooses the coding rate to be
\begin{equation}
    \frac{n}{N}\log(p)=\log\left(\frac{4}{V_{2N}^{2/2N}}\right)+\epsilon,
\end{equation}
for $\epsilon>0$. The volume of the fundamental Voronoi region is lower bounded by
\begin{align}
    \text{Vol}(\mc{V}_{\Lambda})^{2/2N} & \geq \left(\gamma^{2N} \frac{p^{N}}{p^{n}} \sqrt{|d|}^N \right)^{2/2N} \nonumber \\
    &= (\gamma^{2N}p^{N-n}\sqrt{|d|}^N)^{2/2N} \nonumber \\
    &= 2^{-\epsilon} NPV_{2N}^{2/2N}.
\end{align}
Hence, we have
\begin{align}
    \lim_{2N\rightarrow \infty} \mbb{E}_{\Lambda}\left(G_{\Lambda}\right) &=
    \lim_{2N\rightarrow \infty} \mbb{E}_{\Lambda}\left(\frac{\sigma^2(\Lambda)}{\text{Vol}(\mc{V}_{\Lambda})^{2/2N}}\right) \nonumber \\
    &\leq \lim_{2N\rightarrow \infty} \frac{\mbb{E}_{\Lambda}(\sigma^2(\Lambda))}{2^{-\epsilon} 2NP V_{2N}^{2/2N}} \nonumber \\
    &= 2^{\epsilon}\lim_{2N\rightarrow\infty}\frac{1}{2N}V_{2N}^{-2/2N} \nonumber \\
    &= 2^{\epsilon}\frac{1}{2\pi\exp(1)}.
\end{align}
Similar to \cite{ordentlich_erez_simple}, one can then use the above result to show that asymptotically, most of the lattices thus constructed will be good for MSE quantization.

%From Chebyshev's inequality, one has
%\begin{equation}
%    \Pp\left(\sum_{i=1}^W \chi_i=0\right)\leq \frac{\var(\frac{1}{W}\sum_{i=1}^W \chi_i)}{\varepsilon^2},
%\end{equation}
%where the nominator can be further bounded
%\begin{equation}
%    \var\left(\frac{1}{W}\sum_{i=1}^W \chi_i\right)=\frac{1}{W^2}\sum_{i=1}^W\sum_{l=1}^W \textrm{Cov}(\chi_i,\chi_l)\leq\frac{q\varepsilon}{W}.
%\end{equation}
%Substituting

\bibliographystyle{ieeetr}
\bibliography{journal_abbr,bib_alternating_CF}

\end{document}